\newif\ifpublish
\publishfalse

\documentclass[conference]{IEEEtran}
\IEEEoverridecommandlockouts
\usepackage{cite}
\usepackage{amsmath,amssymb,amsfonts,amssymb}
\usepackage[mathscr]{euscript}
\newmuskip\pFqmuskip

\usepackage{algorithmic}
\usepackage{graphicx}
\usepackage{textcomp}
\usepackage{xcolor}
\usepackage{xspace}
\usepackage[inline]{enumitem}
\usepackage{array}
\usepackage{makecell}
\newcolumntype{x}[1]{>{\centering\arraybackslash}p{#1}}

\usepackage{dblfloatfix}
\usepackage[caption=false,font=normalsize,labelfont=sf,textfont=sf]{subfig}
\usepackage{pgfplots}
\usepackage{tikz}
\usepackage{forest}
\usepackage{multirow}
\usetikzlibrary{chains,decorations.pathreplacing,scopes}
\usetikzlibrary{shapes.multipart}
\usetikzlibrary{calligraphy}
\tikzstyle{arrow} = [thick,->,>=stealth]
\def\BibTeX{{\rm B\kern-.05em{\sc i\kern-.025em b}\kern-.08em
    T\kern-.1667em\lower.7ex\hbox{E}\kern-.125emX}}
\usetikzlibrary{positioning, shapes, arrows, quotes}
\usepackage{varwidth}
\usepackage{ragged2e, colortbl} 

\usepackage{algorithm}

\providecommand{\name}{\textnormal{Bonsai}\xspace} 
\providecommand{\bonsai}{\name} 
\providecommand{\seed}{\ensuremath{seed}}

\providecommand{\user}{\ensuremath{{\sf Client}}\xspace} 
\providecommand{\users}{\ensuremath{{\sf Clients}}\xspace} 
\providecommand{\cloud}{\ensuremath{{\sf Cloud}}\xspace} 
\providecommand{\system}{\ensuremath{\mathscr{S}}\xspace} 
\providecommand{\userSet}{\ensuremath{{U}}\xspace} 
\providecommand{\userCount}{\ensuremath{{n}}\xspace} 

\providecommand{\file}{\ensuremath{F}\xspace} 
\providecommand{\base}{\ensuremath{{B}}\xspace} 
\providecommand{\Addendum}{\ensuremath{{A}}\xspace} 
\providecommand{\Change}{\ensuremath{{S}}\xspace} 
\providecommand{\ChangedValues}{\ensuremath{{C}}\xspace} 
\providecommand{\chnvResult}{\ensuremath{{G}}\xspace} 
\providecommand{\sepResult}{\ensuremath{{H}}\xspace} 
\providecommand{\baseset}{\ensuremath{\mathcal{B}}\xspace} 

\providecommand{\id}{\ensuremath{{\sf fid}}\xspace} 
\providecommand{\local}{\ensuremath{L}\xspace} 

\providecommand{\singlefile}{\ensuremath{\mathcal{F}}\xspace} 
\providecommand{\singlesource}{\ensuremath{\mathcal{F}^{\prime}}\xspace} 

\providecommand{\policy}{\ensuremath{{\sf Policy}}\xspace} 
\providecommand{\outsource}{\ensuremath{\file^{\prime}}\xspace} 
\providecommand{\database}{\ensuremath{DB}\xspace} 
\providecommand{\basepointer}{\ensuremath{\mathcal{P}}\xspace}  

\providecommand{\transformation}{\ensuremath{{\sf{trf.Client}}}\xspace}
\providecommand{\cloudtransformation}{\ensuremath{{\sf{trf.Cloud}}}\xspace} 
 
\providecommand{\deletions}{\ensuremath{{\sf{Del}}}\xspace} 
\providecommand{\changev}{\ensuremath{{\sf ChngV}}\xspace} 
\providecommand{\swap}{\ensuremath{{\sf{Swap}}}\xspace}  
\providecommand{\invert}{\ensuremath{{\sf{Inv}}}\xspace} 
\providecommand{\setup}{\ensuremath{{\sf{SetUp}}}\xspace} 

\providecommand{\decompress}{\ensuremath{{\sf{Deco}}}\xspace} 
\providecommand{\upload}{\ensuremath{{\sf{Upload}}}\xspace} 
\providecommand{\get}{\ensuremath{{\sf{Get}}}\xspace} 
\providecommand{\sort}{\ensuremath{{\sf{Sort}}}\xspace} 
\providecommand{\dedup}{\ensuremath{{\sf{Deduplication}}}\xspace} 
\providecommand{\changeVs}{\ensuremath{{\sf{Change Values}}}\xspace} 
\providecommand{\sep}{\ensuremath{{\sf{Seperation}}}\xspace} 
\providecommand{\baseSort}{\ensuremath{{\sf{Base Sort}}}\xspace} 

\providecommand{\Bid}{\ensuremath{{\sf{Bracket ID}}}\xspace} 
\providecommand{\Sid}{\ensuremath{{\sf{Symbol ID}}}\xspace} 
\providecommand{\Vid}{\ensuremath{{\sf{Value ID}}}\xspace} 

\providecommand{\bits}{\ensuremath{k}\xspace}  
\providecommand{\alphabetSize}{\ensuremath{N}\xspace}  

\providecommand{\nodel}{\ensuremath{\alpha}\xspace}  
\providecommand{\nof}{\ensuremath{{\sf{\#}}\file}\xspace}  
\providecommand{\nob}{\ensuremath{{|\baseset|}\xspace}}  

\providecommand{\sorg}{\ensuremath{n_{\file}}\xspace} 
\providecommand{\sbase}{\ensuremath{n_{\file^{\prime}}}\xspace} 
\providecommand{\sizefid}{\ensuremath{s_{\id}}\xspace}  
\providecommand{\sizep}{\ensuremath{s_{p}}\xspace}  
\providecommand{\sseed}{\ensuremath{s_{seed}}\xspace}  

\providecommand{\distance}{\ensuremath{ {\sf I}}\xspace} 
\providecommand{\test}{\ensuremath{t}\xspace} 

\providecommand{\cons}{\ensuremath{c}\xspace} 
\providecommand{\invertbit}{\ensuremath{I}\xspace} 
\providecommand{\sizeforest}{\ensuremath{s_{\baseset}}\xspace} 

\providecommand{\nswap}{\ensuremath{n_{swap}}\xspace} 

\providecommand{\compratio}{\ensuremath{\mathcal{C}}\xspace} 
\providecommand{\ucr}{\ensuremath{\compratio_\user}\xspace} 	
\providecommand{\ccr}{\ensuremath{\compratio_\cloud}\xspace} 	
\providecommand{\unmetric}{\ensuremath{\mathcal{U}}\xspace} 	

\providecommand{\guess}{\ensuremath{q}\xspace} 	
\providecommand{\distribution}{\ensuremath{\mathcal{D}}\xspace} 	
\providecommand{\clouddistribution}{\ensuremath{\mathcal{D^{\prime}}}\xspace} 	
\providecommand{\leakage}{\ensuremath{\mathcal{L}}\xspace} 	

\providecommand{\size}{\ensuremath{{{\sf size}}}}					
\providecommand{\adversary}{\ensuremath{\mathcal{A}}\xspace}		
\providecommand{\Wadversary}{\ensuremath{\mathcal{A}^{\sf week}}\xspace}		
\providecommand{\Sadversary}{\ensuremath{\mathcal{A}^{\sf strong}}\xspace}		
\providecommand{\nopreim}{\ensuremath{m}\xspace}  
\providecommand{\Pset}{\ensuremath{\mathcal{I}}\xspace}  
\providecommand{\alphabet}{\ensuremath{\mathbb{A}}\xspace} 

\usepackage[colorinlistoftodos]{todonotes}

\newtheorem{theorem}{Theorem}

  \tikzset{
  	max width/.style args={#1}{
  		execute at begin node={\begin{varwidth}{#1}},
  			execute at end node={\end{varwidth}}
  	}
}
  
\begin{document}
	\setlength{\extrarowheight}{0.1cm}
	
\title{\name: A Generalized Look at Dual Deduplication
}

\author{\IEEEauthorblockN{Hadi Sehat$^{1}$, Anders Lindskov Kloborg$^{2}$, Christian Mørup$^{2}$, Elena Pagnin$^{2}$, Daniel E. Lucani$^{1}$}
  \IEEEauthorblockA{$^{1}$Agile Cloud Lab, Department of Engineering, DIGIT, Aarhus University, Aarhus, Denmark\\
  	 $^{2}$ Department of Engineering, Aarhus University, Aarhus, Denmark\\
    $^{3}$ Department of Electrical and Information Technology, Lund University, Lund, Sweden\\
\{hadi,daniel.lucani\}@eng.au.dk, \{anderskloborg, Chmorup\}@gmail.com, elena.pagnin@eit.lth.se}
\vspace{-1.5em}
}
\vspace{-1.5em}

\maketitle
\begin{abstract}

Cloud Service Providers (CSPs) offer a vast amount of storage space at competitive prices to cope with the growing demand for digital data storage. Dual deduplication is a recent framework designed to improve data compression on the CSP while keeping clients' data private from the CSP. To achieve this, clients perform lightweight information-theoretic transformations to their data prior to upload. We investigate the effectiveness of dual deduplication, and propose an improvement for the existing state-of-the-art method. We name our proposal Bonsai as it aims at reducing storage fingerprint and improving scalability. In detail, Bonsai achieves (1) significant reduction in client storage, (2) reduction in total required storage (client + CSP), and (3) reducing the deduplication time on the CSP. Our experiments show that Bonsai achieves compression rates of 68\% on the cloud and 5\% on the client, while allowing the cloud to identify deduplications in a time-efficient manner. We also show that combining our method with universal compressors in the cloud, e.g., Brotli, can yield better overall compression on the data compared to only applying the universal compressor or plain Bonsai. Finally, we show that Bonsai and its variants provide sufficient privacy against an honest-but-curious CPS that knows the distribution of the Clients' original data.

\end{abstract}


\section{Introduction}
\label{intro}

Due to emerging applications that produce massive amounts of data,
many systems choose to use Cloud Service Providers (CSPs) to outsource the storage of their data.
In order to deliver a cost-effective service to their clients, CSPs tend to use compression techniques to reduce the fingerprint of the data.
This includes a number of file compression techniques, e.g., DEFLATE \cite{deutsch1996rfc1951}, BZIP2 \cite{gilchrist2004parallel}, Brotli \cite{alakuijala2018brotli},
as well as delta encoding approaches, e.g., VCDIFF \cite{korn2002vcdiff},
which focus on individual chunks of data or files for compression.
As an alternative, data deduplication~\cite{Jin2009}
removes duplicate data across different files and even different users in the system.
Data deduplication stores a copy of each unique chunk and represents files with pointers to these unique chunks.
In a nutshell, it avoids storing duplicates \cite{xia2016comprehensive}.
In most practical storage systems, both deduplication and compression are used
for a more efficient reduction of the fingerprint of the data \cite{constantinescu2011mixing}.

Recently, generalized deduplication \cite{rasmus} has been proposed as an improvement to data deduplication.
In this case, the storage system is not only capable of removing exact duplicates,
but also chunks of files that are close to each other according to a certain metric~\cite{Vestergaard2019a}.
At its core, generalized deduplication
in a cloud setting works as follows.
The CSP
 maintains a set of bases. 
For each received chunk of file, 
if the chunk is already present in the set of bases
the CSP deduplicates it 
by storing a file identifier and a pointer representing the base.
Otherwise,
the CSP looks for a base similar to the given chunk,
i.e., a chunk in the set of bases that has a short distance to the received chunk, e.g., according to the
Hamming distance~\cite{Vestergaard2019a} or the Swap distance~\cite{yggdrasil}.
If such a base is found,
the CSP deduplicates this record in a generalized way
by storing a file identifier, a pointer representing the closest base, and a string containing the difference between the base and the chunk. 
Finally, if no similar bases are present in the set of stored bases,
the CSP considers the chunk as a new base to add to the list of bases.

A major drawback of allowing the CSP to carry out (generalized) deduplication
is that the CSP must be provided with the actual data.
This is not desirable in applications where clients may upload sensitive information or are concerned about data privacy.
A naive solution would be to encrypt the data and upload only ciphertexts.
This approach, however, drastically reduces the deduplication capability on the CSP, as secure encryption implies little or no correlation among ciphertexts.
An alternative approach is to use
convergence encryption~\cite{bellare2013message},
to maintain a balance between privacy and compression.
The main weakness of convergence encryption
is that a nosy CSP or even a malicious client can attack the system to discover which chunks have been stored previously in the system,
exposing sensitive information to the malicious party.
Moreover, the CSP can obtain exact information about the files of each user given sufficient time and resources, by breaking the encryption of the outsourced data~\cite{akhila2016study}.

The work by Vestergaard et al.~\cite{lucani2020secure} proposes an alternative mechanism to convergence encryption, which is called Multi-Key revealing Encryption (MKRE).
In~\cite{lucani2020secure} chunks are transformed before upload so that any two similar chunks yield outsourced data that are close to each other.
Although  this method is shown to be more robust against attempts to break the encryption,
its security is proven only in the programmable random oracle model,
which is not realistic as it cannot be faithfully implemented by hash functions.

Recently, the work by Sehat et al.~\cite{yggdrasil} bypasses the programmable random oracle challenge and
introduces a unique privacy-aware deduplication mechanism for multi-client environments called \emph{dual deduplication}.
In dual deduplication, clients perform
information-theoretic transformations on their data
before uploading it to the CSP.
These transformations 
aim at puncturing 
the data
to simultaneously increase deduplication capabilities on the CSP,
and creating uncertainty for the CSP about the possible pre-images
of the uploaded chunk.
From a high level perspective,
to the eye of a CSP or an attacker,
the dual deduplication system
proposed in~\cite{yggdrasil} mimics a \emph{deletion channel} from information theory \cite{mitzenmacher2009}.  
In particular it is known that
retrieving the original data
after transmission through the deletion channel
 is a hard problem.
predicting how input data will be modified by the channel is hard as well~\cite{mitzenmacher2009}.
While these properties are undesirable to establish a reliable communication between two parties,
dual deduplication uses them as key features for proving the privacy of the system.
In this work, we build on the dual deduplication idea introduced in~\cite{yggdrasil},
and mitigate the three major drawbacks of the original construction.
First, clients in~\cite{yggdrasil} need to store a significant portion of their data locally.
This is contrary to the intent of outsourcing data to a CSP and minimizing local storage.
Second, the CSP in~\cite{yggdrasil} must perform a brute force search to find the matches for each uploaded chunk to achieve the best compression rate.
Third, the system in \cite{yggdrasil} is proven private only as long as
the CSP lacks information about the probability distribution of the client's original data.
This assumption does not hold for databases that are very homogeneous,
or with predictable data such as time-series data or e-mail texts.

The main goal of our paper is to face the aforementioned drawbacks
and provide concrete, efficient solutions to address these issues.
To this end, we propose \emph{Bonsai}, a fully fledged, yet scaled down (in terms of storage requirements and complexity) alternative to the Yggdrasil approach proposed in \cite{yggdrasil}.
The name \name is chosen to maintain a connection to and provide figurative comparison with the original dual deduplication method Yggdrasil, named after the tree of life in Norse mythology.
Moreover, in \name the CSP transforms and stores the received data in a tree structure that resembles a bonsai, which is
a small-size version of a tree.

\subsection{Our Contribution}

We propose a new transformation mechanism for dual deduplication
that improves the \emph{scrambling} of outsourced data,
	and achieves better privacy guarantees than~\cite{yggdrasil}.
	Compared to~\cite{yggdrasil}, our method requires less storage on the client side and maintains
	good compression rates on the CSP.
	This is done by employing a Pseudo-Random Number Generator (PRNG)
	 to determine the positions of the symbols to be deleted.
	 This change removes the need to store the positions of deleted elements on the client side, thus effectively reducing the storage requirements from 18\% in \cite{yggdrasil}, to only 5\% in \name.

We present an innovative deduplication mechanism that lets the CSP
	identify possible deduplications efficiently (instead of bruteforcing as in \cite{yggdrasil}).
	Our technique stores bases on the CSP in a forest data structure, which
	reduces the computational cost of finding deduplicates and adding new bases as well as reducing the storage cost for the bases.
	In addition, we improve the compression potential on the CSP using an approach inspired by Huffman Codes and replacing the symbols with lower probability distribution.
	Our method achieves a competitive compression rate of 68\% in the cloud, while allowing deduplication and maintaining privacy guarantees.

We prove our proposed method to be privacy-aware, in the sense that
an honest-but-curious CSP faces a high degree of uncertainty when guessing what the original data of the clients is.
In contrast to~\cite{yggdrasil}, this claim holds even assuming the CSP has knowledge of the distribution of clients' original data,	as well as of the number of performed deletions.

\begin{figure}[!t]
	\centering

	\begin{tikzpicture}
		[scale=0.65]
		\begin{axis}[
			xlabel= $chunksize \,(Bytes)$,
			ylabel=$Compression\; rate$,
			xlabel style={font=\large},
			xtick = {6,8,10,12,14,16},
			xticklabels = {$2^{6}$,$2^{8}$, $2^{10}$, $2^{12}$, $2^{14}$, $2^{16}$},
			ymin=0,
			ymax = 1.3,
			grid=major,
			legend pos= outer north east]

			\addplot[mark= +, smooth ,black, thick] plot coordinates {
				(5, 0.84)
				(6, 0.88)
				(7, 0.94)
				(8, 0.98)
				(9,1.02)
				(10,1.05)
				(11,1.09)
				(12, 1.12)
				(13, 1.14)
				(14,1.14)
				(15,1.14)
				(16,1.14)
			};
			\addlegendentry{SecDedup\cite{zhang2020secdedup}}

			\addplot[mark = o, smooth, red, thick] plot coordinates {
				(5, 0.54)
				(6, 0.56)
				(7, 0.58)
				(8, 0.64)
				(9,0.69)
				(10,0.74)
				(11,0.78)
				(12, 0.83)
				(13, 0.87)
				(14, 0.91)
				(15,0.94)
				(16,0.97)
			};
			\addlegendentry{\name}

			\addplot[mark=x, blue, thick] plot coordinates {
				(5, 0.54)
				(6, 0.58)
				(7, 0.62)
				(8, 0.69)
				(9,0.76)
				(10,0.82)
				(11,0.87)
				(12, 0.96)
				(13, 1.03)
				(14, 1.09)
				(15, 1.17)
				(16, 1.26)
			};
			\addlegendentry{Yggdrasil \cite{yggdrasil}}

			\addplot[mark=diamond, smooth,teal, thick] plot coordinates {
				(5, 0.32)
				(6, 0.33)
				(7, 0.35)
				(8, 0.37)
				(9, 0.39)
				(10, 0.42)
				(11, 0.45)
				(12, 0.48)
				(13, 0.52)
				(14,0.55)
				(15,0.57)
				(16,0.59)
			};
			\addlegendentry{Brotli\cite{alakuijala2018brotli}}

			\addplot[mark=oplus, smooth,violet, thick] plot coordinates {
				(5, 0.27)
				(6, 0.25)
				(7, 0.25)
				(8, 0.26)
				(9,0.28)
				(10,0.30)
				(11, 0.32)
				(12, 0.34)
				(13, 0.36)
				(14,0.37)
				(15,0.38)
				(16, 0.39)
			};
			\addlegendentry{\name + Brotli}

		\addplot[mark=pentagon, dashed, smooth, brown, thick, mark options={solid}] plot coordinates {
			(5, 0.23)
			(6, 0.22)
			(7, 0.24)
			(8, 0.26)
			(9,0.28)
			(10, 0.30)
			(11, 0.32)
			(12, 0.34)
			(13, 0.36)
			(14, 0.37)
			(15, 0.38)
			(16, 0.39)
		};
		\addlegendentry{SecDedup +\name + Brotli}

\end{axis}
\end{tikzpicture}
\caption{Total compression rates for different chunk sizes using \name (this work), Yggdrasil~\cite{yggdrasil}, SecDedup \cite{zhang2020secdedup}, Brotli~\cite{alakuijala2018brotli}, and combination of these techniques (lower is better). }
\label{fig:intro}
\end{figure}
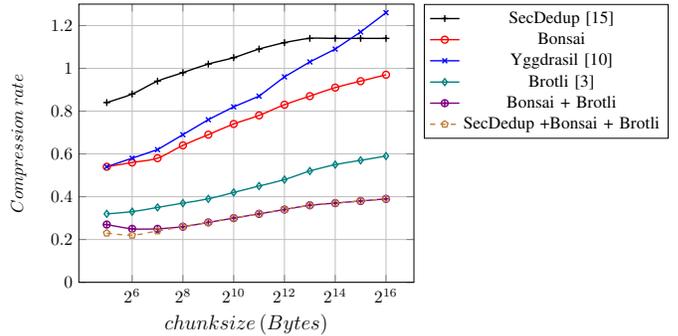

Finally, to showcase the performance impact of \name on a cloud storage system, we used a DVI dataset of 10~GB and collected simulations for compression ratios achieved by \name, Yggrdasil~\cite{yggdrasil}, data deduplication (DD)~\cite{zhang2020secdedup}, and Brotli~\cite{alakuijala2018brotli}.
The comparison is illustrated in Fig.~\ref{fig:intro}, where plots display the ratios between the total storage required after compression (considering both client and CSP) and the size of the actual data.
A compression rate of 1 indicates no compression and values above 1 expand the size of the data.
The lower the compression rate, the less storage is needed for each original bit (i.e., better compression).
 As Fig~\ref{fig:intro} suggests,
 \name has better compression rate than its predecessor Yggrasil \cite{yggdrasil}, especially on larger chunk sizes.
Brotli outperforms \name in compression rate,
which is expected as \name is mostly a deduplication technique, which tend to have lower compression rates than universal compressors.
However, Brotli provides no privacy guarantees, in contrast to \name.
Interestingly, combining \name with Brotli (i.e., using Brotli to compress the result of the deduplication process) yields much better compression than each individual approach.
 This suggests that \name creates highly compressible data, even when the original data is homogenous.
Finally, applying data deduplication to find repeated blocks of data before applying \name and Brotli further increases the compression gain for low chunk size.
Intuitively, this is due to the fact that \name adds metadata to find matches across \emph{different} data blocks, which is unnecessary for deduplicating identical chunks.
The gains are seen for smaller chunk sizes, 
disappearing for larger chunk sizes due to the fact that the probability of finding an exact duplicates is reduced as the size of the chunks increases.

%

The outline of this paper is as follows.
In Section~\ref{sec:RelWork}, we describe the related work on (generalised) deduplication and privacy in systems using deduplication.
In Section~\ref{sec:model}, we present our system model.
In Section~\ref{sec:contribution}, we propose \name and clearly describe its core algorithms and processing steps. Section~\ref{sec:analysis} includes the analysis of the system in terms of privacy, compression rate and computation cost.
Section~\ref{sec:experiments} includes our simulation results and discussion about the
performance of \name.
Finally, we conclude in Section~\ref{sec:conclusion}.

\section{Related Work}\label{sec:RelWork}

Data deduplication is a powerful way to reduce the storage footprint of highly correlated data \cite{Jin2009,meyer2012study}.
 Delta deduplication falls into the category of redundant data reduction techniques, which also include Delta Compression techniques such as Xdelta~\cite{macdonald2000file}.
 The big difference between the two methods is the facts that Delta encoding techniques use copy/insert instructions to record the ``difference'' between two files
 on string-level, while the data Deduplication techniques remove the ``redundancy'' between files, either on file-level~\cite{bolosky2000single} or chunk-level~\cite{quinlan2002venti}.
 The theoretical bounds of deduplication and generalized deduplication are discussed extensively in \cite{niesen2019information, Vestergaard2019a, lou2020data}.
 Implementations of generalized deduplication have been proposed for file systems~\cite{MinervaFS} and CSPs~\cite{Alexandria}.

 The introduction of deduplication in the early 2000s sparked many research directions aiming at improving or enhancing the compression method.
 Problems and proposed solutions can summarized into three major categories.
 \begin{itemize}
 	\item \textbf{Increasing performance and throughput:} This line of research includes, but is not limited to optimization of deduplication~\cite{zhang2020secdedup},
 	acceleration of computational tasks~\cite{ren2021accelerating},
 	reduction of the access time to metadata~\cite{zhu2008avoiding},
 	 and improvements of the chunking procedure based on the content~\cite{yu2015leap, zhou2013hysteresis}.
 	\item \textbf{Reducing the storage requirement:} These solutions focus on reducing the fingerprint of the data.
 	Some of the major research outputs in this category deal with reducing the overhead of the metadata~\cite{li2019metadedup},
 	improving the efficiency of deduplication in the cloud and reducing the size of fingerprints~\cite{bhagwat2009extreme},
 	and reducing the overhead of the data by establishing a communication
 	between the cloud and the clients \cite{vestergaard2020cider, pooranian2018rare, yu2018privacy}.
 	\item \textbf{Improving the privacy and security of the clients' data:} Privacy issues arise in Data Deduplication as all the data from the clients are deduplicated together, leaving the storage system vulnerable to leaking the information of the data. As this category is the closest line of research to the solution proposed in this paper, we take a closer look at this category.
 \end{itemize}

For privacy-aware data owners (clients), 
the CSP should be able to provide storage solutions without needing access to 
the content of their data. 
If conventional encryption is used to protect clients' data, the compression potential on the cloud severely worsens~\cite{akhila2016study}.
As an alternative, Message Leaked Encryption (MLE)~\cite{douceur2002reclaiming} has been proposed, which allows one to encrypt data securely, while preserving the deduplication potential.
The most widely used MLE technique in deduplication techniques is called Convergence Encryption (CE)~\cite{whiting1998system}.
However, CE has been shown to be exploitable by brute-force~\cite{almrezeq2021enhanced},
and is vulnerable to frequency attacks~\cite{li2020information}.
The proposed adjustments to improve CE, such as DUPLESS~\cite{keelveedhi2013dupless}, and Duan's scheme~\cite{duan2014distributed}, require a trusted cloud or a third-party dealer to ensure the privacy of the data.
Alternative approaches rely on the Blockchain technology~\cite{zhang2019blockchain}, on gateways~\cite{leontiadis2018secure},
sacrifice computational efficiency~\cite{zhao2017updatable},
or generate large overhead for the clients~\cite{liu2015secure}. 
More recently, Multi-Key Revealing Encryption~\cite{lucani2020secure} has been proposed to ensure privacy of the clients' data while maintaining deduplication capabilities. This solution uses programmable random oracles which are not instantiatable in practice. 

There have been surveys that describe the privacy solutions in more depth~\cite{zhang2020secure,shin2017survey}.
We also note that~\cite{xia2016comprehensive} provides a general survey on deduplication with a comprehensive study of the state-of-the-art.

\section{System Model and Performance Metrics}\label{sec:model}

In this section, we describe our system model, the type of data handled by the system, the attacker model and, finally, the performance metrics used to evaluate \name.

\subsection{System Model}

We adopt the dual deduplication system introduced in \cite{yggdrasil}. 
Concretely, our system \system consists of a number of independent users, referred to as \users, and a single CSP, referred to as \cloud. 
The \users' goal is to maximize the data they outsource to the \cloud\ while retaining some level of privacy on each record; the \cloud's goal is to minimize the space required to store the \users' outsourced data. 
The system \system allows the \users to locally process their data, by applying a transformation \transformation on their data before upload
 and storing the information about the transformations 
 locally.
Note that the \cloud has no access to the information stored locally by any \user. 
This fact is crucial to argue the privacy of \name. 
Finally, we assume that the communication channel between the \users and \cloud is authenticated and error-free.

\subsection{Data Type}
The data handled by \cloud and \users in \system are files  
\file represented as strings of \sorg symbols,
where each symbol 
is a value
consisting of \bits bits.
i.e., our alphabet is $\alphabet=\{0,1\}^\bits$ 
and files are $\file\in \alphabet^{\sorg}=\{0,1\}^{\bits \cdot \sorg}$.
We denote the size of the alphabet as $\alphabetSize = |\alphabet| = 2^{k}-1$.

Recall that each \user performs some transformations \transformation on their files \file prior to upload.
In what follows, we denote the output of \transformation as an outsource-local pair $(\outsource, \local)\gets\transformation(\file)$, where \outsource, called outsource is the data
 outsourced to the \cloud; and \local is the secret piece of information stored locally by \user.
 We call this secret information client-side deviation. 
Note that both outsource and client-side deviation are strictly shorter than the original file; namely the length of outsource $|\outsource| = \sbase < \sorg$ and the length of the client-side deviation $|\local| < \sorg$.

\subsection{Attacker Model}\label{sec:sec_model}
Our adversary \adversary is a honest-but-curious \cloud.  
Concretely, \adversary has access to the whole data stored in the \cloud, 
i.e., any record \outsource ever outsourced by any \user. 
The goal of the adversary is to breach \user's privacy by retrieving their original data \file. 
i.e., to identify the complete string \file\ from which the \user derived the outsource \outsource. 
We recall that \adversary has no access to the client-side deviation \local\ stored by the \user locally. 
In this work we consider two types of adversaries:

$\bullet$ A \emph{weak} adversary, \Wadversary, does not have any information on the distribution of clients' original data \file. 
This weak adversary emulates settings where clients upload unpredictable data, e.g., very diverse types of files, or uncorrelated information.

$\bullet$ A \emph{strong} adversary, \Sadversary, knows the distribution \distribution of the original files, i.e., the probability distribution with which a specific string \file is produced by a \user. 
Notably, \Sadversary has knowledge of the correlation between the symbols in \file. 

As expected, the strong adversary is more realistic than the weak one.
While it is reasonable to assume that in real systems the CSP may have some information about the files generated by the clients; 
knowing the exact probability distribution for any possible file \file is a requirement hard to meet. 
Hence, we believe a real-world attacker lies somewhere in-between the two types of adversaries we consider.

Our protocol, \bonsai, makes minimal use of cryptographic tools. Namely, the only cryptographic primitive required is a PRNG (Pseudo-Random Number Generator) function, run by the \users within the \transformation process. All other transformations employed in \bonsai are information theoretic. 
To further strengthen our security outcomes, for each type of adversary, we consider two potential scenarios:

$\bullet$ \emph{Case 1}: \adversary has negligible probability of breaking the PRNG (this correspond to a classical adversary that is computationally bounded and runs in probabilistic polynomial time).

$\bullet$ \emph{Case 2}: \adversary can break the PRNG (this corresponds to quantum adversary that is computationally unbounded and equipped with a quantum computer, or an adversary that knows the seed input to the PRNG).

\noindent
We remark that, in \bonsai the PRNG is employed to provide better compression on the \user side. Therefore, it is possible to obtain a complete information theoretic secure version of protocol by replacing the use of the PRNG with a truly random function, though this impacts negatively the performance of the system
by reducing the compression potential on the \cloud.

Finally we argue that assuming \adversary is not fully malicious is needed to guarantee the correctness of the overall system. For example, if the \cloud tampers with outsourced items, \users may reconstruct incorrect data and the service loses reliability. 
We consider that in the specific setting of outsourced storage, an honest-but-curious attacker is more realistic than a malicious one. 
For this reason, we leave the analysis of the implications of having a malicious \cloud or a malicious \user as future work.

\subsection{Compression Ratios}
The main goal of \name is to 
reduce the fingerprint of the data via dual deduplication 
(i.e., compressing data on both the \user and the \cloud sides). 
Therefore, it is natural to define performance metrics that measure the compression ratio of the system by looking at the \user, the \cloud, and the overall system perspective. 
We employ the performance metrics introduced by Yggdrasil, the state-of-the-art work on dual deduplication \cite{yggdrasil}.

In what follows, \database denotes the set of all the data of all the \users ;  
and  $\database_{i}$ denotes the set of the data of a single \user $i$. 
Both \database and $\database_{i}$ contain a number of files \file.
 We assume that our system \system has a total of \userCount \users, 
 which are enumerated in a set $\userSet =\{ \user_{1}, ...., \user_{\userCount} \}$.
In order to formally define the compression rate of \name in different parties (\user and \cloud),
we use a function size(party, D),
which we define as follows.

\begin{description}
	\item[size(\sf{party}, D):] 
	
	On the input of $\sf{party}\in \{\cloud,\user\}$ and some data D,
	this function outputs 
	the total number of bits required to store D in the given party.
\end{description}

Equipped with this notation we can define the \user Compression ratio as :
	\begin{equation}\label{eq:ucr}
		\ucr = \frac{\size(\user_{i}, \database_{i})}{|\database_{i}|},
	\end{equation}
	where $|\database_{i}|$ denotes the total size of the data in the posession of $\user_{i}$.

	As the \cloud stores data from all the clients,
	we calculate the total compression rate of the \cloud with respect to 
	\database, therefore, the \cloud compression rate is defined as:
	\begin{equation}\label{eq:ccr}
		\ccr = \frac{\size(\cloud, \database)}{|\database|}.
	\end{equation}
 	
Another interesting compression metric is the total compression rate, which indicates the total storage required by all the parties (\users and the \cloud) to store the whole 
database 
\database compared to the size of the original \database. This metric allows us to compare the performance of \name 
with other compression methods that only store the data on the \cloud.
Intuitively, the total required storage in \name is equal to the summation of the required storage in all \users and the required storage on the \cloud, therefore, 
the total compression rate is:
%
	\begin{equation}\label{eq:gcr}
		\footnotesize
		\compratio  = \frac{\size(\cloud, \database)+\sum\limits_{i=1}^{\userCount}\size(\user, \database_{i})}{|\database|}.
	\end{equation}

\subsection{Privacy Metrics}

We estimate the privacy level of a dual deduplication system \system in terms of the uncertainty metric \unmetric and the leakage \leakage.
%
The uncertainty metric shows the amount of unpredictability
that an adversary \adversary faces when 
it attempts to guess the value of a given record \file, having seen only its outsourced version \outsource. 
We acknowledge that the adversary may have some knowledge about the file \file prior to receiving the \outsource,
therefore, after receiving the \outsource, the uncertainty of the adversary drops as it gains more information about the file \file.
The leakage \leakage measures this drop in the uncertainty of the adversary \adversary after receiving \outsource, i.e., the amount of information about \file leaked to \adversary after receiving \outsource. 

We use the notation of Shannon's entropy \cite{cover1999elements} to calculate the uncertainty metric and leakage for our adversaries.
We assume that \adversary uses its knowledge to generate the list of all possible preimages $\Pset\gets\adversary(\distribution, \file')$ of a given base \outsource. It is clear that the original file belongs in this list, $\file\in\Pset$. Then the adversary tries to predict the real value of \file using the preimages in \Pset and the probability that each one of them is the real \file.
The uncertainty that the attacker faces to guess the original file \file is equal to the entropy of the set \Pset. Therefore, we define uncertainty metric as follows.
\begin{description}
	\item[Uncertainty Metric:] 
	\begin{equation}\label{eq:uncertainty}
	\unmetric(\file) = H(\file|\outsource),
	\end{equation}
\end{description}
where $H(\file|\outsource)$ is the Shannon's entropy of \file given \outsource.
Based on our definition, the uncertainty prior to receiving \outsource is equal to $H(\file)$. Therefore, the leakage of information that occurs when sending \outsource is calculated as follows.
\begin{description}
	\item[Leakage:]
\begin{equation}\label{eq:leakage}
	\leakage = \frac{H(\file) - H(\file|\outsource)}{H(\file)}. 
\end{equation}
\end{description}

Note that the Shannon's entropy of file, i.e., $H(F)$ is calculated in the viewpoint 
of the adversary.
This means that if we consider the weak adversary, the value for $H(F)$ is equal to the Shannon's entropy for 
a uniformly random probability distribution and not the actual value of the Shannon's entropy
based on the statistical properties of the database \database. 
%
%
%
%

\vspace{0.1cm}
\section{The \name Protocol}\label{sec:contribution}

\begin{figure*}[t]
	\centering
	\begin{tikzpicture}[
		node distance=0pt,
		start chain = A, 
		X/.style = {rectangle, draw,
			minimum width=3ex, minimum height=3.5ex,
			outer sep=0pt, on chain},
		B/.style = {decorate,
			decoration={brace, amplitude=5pt,
				pre=moveto,pre length=1pt,post=moveto,post length=1pt,
				raise=1mm,
				#1}, 
			thick},
		B/.default=mirror, 
		]

		\foreach \i in {4,10,1,8,9,7}
		\node[X,right] {\i};

		\node[X,text=red,right] {1};

		\foreach \i in {2,2,
			12}
		\node[X,right] {\i};

		\node (end)[X, right] {15};
		\node (begin)[X, right = 1.3cm] {4};

		\foreach \i in {10,1,8,9,7,2,2,
			12,15}
		\node[X, right] {\i};


		\node[X,text=blue, right = 1cm] {6};
		\node[X, text=red, right] {1};

		\draw [arrow] (end)  -- (begin);
		\node (B1) [inner sep=1pt,above=of A-7.north west] {Input};
		\node (B2) [inner sep=1pt,above=of A-17.north west] {Output};
		\node (B3) [inner sep=1pt,above=of A-23.north west] {Deviation ($\local$)};
		\foreach \x [count=\i] in {0, ...,6}
		\node [below=of A-\i]{\textcolor{black!50}{\x}};
		\node [below=of A-7]{\textcolor{blue}{6}};
		\foreach \x [count=\i] in {7, ...,10}{
		\pgfmathtruncatemacro{\y}{\x+1};
		\node [below=of A-\y]{\textcolor{black!50}{\x}};};
		\node[below=of A-1, xshift=-30pt]{\textcolor{black!50}{$position$}};
		\foreach \x  in {0, ...,9}
		\pgfmathsetmacro{\i}{\x+12}
		\node [below=of A-\i,xshift=-5pt, yshift=-6pt]{\textcolor{black!50}{\x}};
		\node [below=of A-22]{\textcolor{blue!50}{{$P$}}};
		\node [below=of A-23]{\textcolor{red!50}{$V$}};
		\node[left=of A-1, xshift = -8pt]{\textcolor{black}{$\deletions:$}};

	\end{tikzpicture}
	\caption{An illustration of \deletions, where \bits=3. In this example the symbol with value 1 from position 6 is deleted from \file.}
	\label{fig:Del}
\end{figure*}
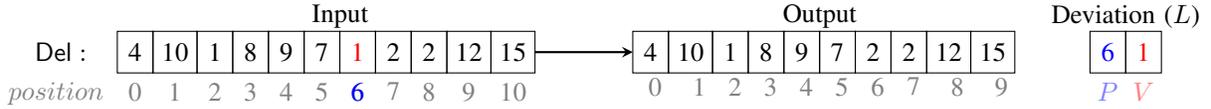

We begin by providing an overview of \name followed by a detailed description of the actions performed in the \cloud and the \user.
We conclude with a detailed explanation of the whole the protocol.

\subsection{\name in a nutshell}\label{sec:transformations}

We employ a similar system model as~\cite{yggdrasil}, namely, a single \cloud storage provider connected to multiple \users. \users wish to outsource their data to the \cloud in a privacy-preserving manner,
while the \cloud wishes to reduce the storage footprint of the received outsourced data.

To achieve privacy-preserving dual deduplication, each \user applies a transformation $\transformation()$ to the original file \file to generate the pair $(\outsource, \local)\gets\transformation(\file)$.
Intuitively, \outsource is the record \user outsources to \cloud,
and this should leak as little information about the original file \file as possible (to preserve privacy).
\local is called the client-side deviation and condenses instructions on how to reconstruct \file from \outsource.
The \transformation entails a number of manipulations as shown in Section \ref{sec:client} and Fig.~\ref{fig:client}.
For the sake of simplicity, we describe only the main manipulation in \transformation, i.e., \deletions (deletion), in more detail.
As the name suggests, this transformation removes  symbols from \file. The outcome is a sub-string \outsource of the original file \file.
To build intuition, \deletions acts as a deletion channel. Thus, it simultaneously performs two tasks
(1) it reduces the size of the outsourced record, 
and
(2) it makes it harder for the \cloud (and thus an adversary \adversary) to guess the exact original data \file from \outsource.
To enable the reconstruction of \file from \outsource,
\deletions returns an additional output that contains the list of deleted positions (in the form of pointers) and the corresponding deleted values (in the form of alphabet symbols)
in form of the client-side deviation \local.
With \local at hand, the \user can efficiently reconstruct \file from \outsource.
Fig.~\ref{fig:Del} illustrates an example of how \deletions works.
In this example, the symbol `1' from position `6' is deleted.
The local deviation contains two values: `6' is a pointer indicating the position of the deleted symbol, and `1' is the deleted symbol.

 In \name, we employ a PRNG function to select the positions where we perform deletions.
 Concretely, the \user selects a pre-determined number of seeds 
 for each chunk, and interprets the output of the PRNG
 of each seed
 as a sequence of positions on which to apply \deletions in the chunk. 
 The use of seeds instead of storing positions alongside values, as done in~\cite{yggdrasil}, significantly reduces the storage requirement on the \user.
 This means that all positions of deleted elements in \local can be stored by a single seed.
 Further details on how to combine the PRNG and \deletions, and the other manipulations involved in \transformation, are given in Section~\ref{sec:client}.
 We also analyze how the use of a PRNG impacts security in Section~\ref{sec:security_proofs}.

\bonsai employs a novel way to perform generalized deduplication on the \cloud side.
Upon receiving an outsource \outsource, instead of looking for similar bases according to
classical distance metrics, e.g., Hamming ~\cite{Vestergaard2019a}
and Swap~\cite{yggdrasil},
the \cloud in \name processes \outsource
to generate a sorted base \base and a set of strings, called \emph{cloud-side deviation}.
After this transformation, the \cloud
applies deduplication on \base.
We call this \cloud-side process \cloudtransformation.
Its aim is to identify an alternative representation of \outsource that 
deduplicates with higher probability.

The manipulations in \cloudtransformation are: grouping, Huffman coding and sorting as detailed in Section~\ref{sec:cloud}.
To give an intuition, 
the \cloud in \name identifies each symbol with three values.
We denote these values by \Bid,
\Sid, and \Vid. 
These three values uniquely identify eqach symbol
(more details in section~\ref{sec:cloud})
.
Upon receiving an \outsource, the \cloud
generates a string containing the \Bid of
for all symbols in \outsource,
and sorts it in ascending order.
This string is called base and
is denoted by \base.
As the base \base is sorted,
the \cloud stores the \base of all of the received \outsource in a forest structure (called \baseset),
to help find duplicates in time $O(\log \nob)$,
where $\nob$ is the number of bases in \baseset.
Using this forest structure, instead of directly storing the \base
for each \outsource,
the \cloud assigns a pointer to \base in \baseset,
and uses this pointer to store each \outsource.
The \cloud needs to store 
the necessary information 
to reconstruct \outsource from \base
upon \user's request.
This information is stored 
in what we call cloud-side deviation.
Cloud-side deviation contains of three strings: \Change, \Addendum, \ChangedValues. 
In detail, 
\Change contains information about the position of the symbols
in the original outsource \outsource; 
\Addendum contains the \Sid of the symbols; and 
\ChangedValues contains the \Vid of the symbols.
\Change is created by using an algorithm inspired by 
Cycle Sort~\cite{bartle2000introduction}
to identify the required swaps
needed to sort \base.
This algorithm
is detailed in Algorithm \ref{alg:cap}, in Section~\ref{sec:cloud}.
In order to reduce the size of the
\Sid and \Vid for the symbols,
and ultimately having a bigger compression gain in the \cloud,
the \cloud uses Huffman coding to represent the \Sid and \Vid of the symbols
in each Bracket. We discuss the choice of Huffman coding  and its implications in Section \ref{sec:cloud}.
\vspace{-0.1cm}
\subsection{\policy and protocols between \cloud and \user}

The communication between \cloud and \user consists of two algorithms: \upload and \get.
With \upload the \user sends \outsource to the \cloud;
while with \get the \user requests back the uploaded data from the \cloud.
In order to ensure consistency, the \user uses a file identifier, \id, for each outsource \outsource it uploads to \cloud. The identifier \id is sent to the \cloud alongside the \outsource and is used by \get as a reference to \outsource. 
Since we assume \cloud not to be malicious, \id does not serve as an integrity tool.  Employing a cryptographic secure hash function would, however, mitigate some attack scenarios in the presence of a malicious cloud, or provide secure file sharing among \users~\cite{sehat2022bifrost}.

In \name, the \cloud performs deduplication to optimize space and reduce the storage requirements.
Therefore, the compression potential increases if the received outsource \outsource are similar to each other.
This is 
 achieved by enforcing a certain probability distribution on the uploaded records.
In order to ensure that the \cloud receives \emph{deduplication friendly} data,
\name adopts policies on the probability distribution of symbols in \outsource. 
Policies are publicly available to \users and should be fetched before \upload and
taken  into consideration when running \transformation.
 In detail, $\policy = [\clouddistribution, \sbase]$
%
where \sbase is the expected size of \outsource and \clouddistribution is defined an array of values $\clouddistribution = [p^{\prime}_{0},\ldots,p^{\prime}_{\alphabetSize-1}]$,
and $p^{\prime}_{i}$ indicates the probability of symbol $i$ in the \cloud.
This policy is calculated by the \cloud from the set of stored bases
\baseset.
This increases the potential for 
deduplication between outsourced data 
by creating \outsource that are
similar to each other,
leading to higher deduplication rates 
and lower storage requirement in the \cloud 
and ultimately
higher compression gain.
The \policy may include more detailed information to further increase compression capability of \cloud or for other use cases such as ensuring reliability.
We leave other information that can be included in \policy and the effect of such information on performance and compression potential of \name to future work.

\subsection{The \user of \name}\label{sec:client}


\begin{figure*}[t]
	\centering
	\begin{tikzpicture}[
		node distance=0pt,
		X/.style = {align=left, rectangle, draw,
			minimum width=2.5ex, max width=2.5ex, minimum height=2.8ex,
			outer sep=0pt},
		Y/.style = {align=left, rectangle, draw,
			minimum width=3.5ex, minimum height=3.5ex,
			outer sep=0pt},
		]


		{
			[start chain =A],
			\node[X,right, on chain = A] at (-2,-1.7) {\footnotesize 4};
			\node[X,right, on chain = A] {\footnotesize 10};
			\node[X,right, on chain = A] {\footnotesize \textcolor{red!80}{1}};
			\foreach \i in {8,9}
			\node[X,right, on chain = A] {\footnotesize \i};
			\node[X,right, on chain = A] {\footnotesize \textcolor{red!30}{7}};
			\foreach \i in {1,2}
			\node[X,right, on chain = A] {\footnotesize \i};
			\node[X,right, on chain = A] {\footnotesize \textcolor{red!60}{2}};
			\node[X,right, on chain = A] {\footnotesize 12};
			\node(endA) [X, right, on chain = A] {\footnotesize 15};
		}
		\foreach \x [count=\i] in {0,1}
		\node [below=of A-\i]{\textcolor{black!50}{\tiny \x}};
		\node [below=of A-3]{\tiny \textcolor{blue!80}{2}};
		\foreach \x [count=\i] in {3,4}
		\pgfmathsetmacro{\i}{\x+3}
		\node [below=of A-\i,xshift=-28pt, yshift=-6pt]{\textcolor{black!50}{\tiny \x}};
		\node [below=of A-6]{\tiny \textcolor{blue!30}{5}};
		\node [below=of A-7]{\tiny \textcolor{black!50}{6}};
		\node [below=of A-8]{\tiny \textcolor{black!50}{7}};
		\node [below=of A-9]{\tiny \textcolor{blue!60}{8}};
		\node [below=of A-10]{\tiny \textcolor{black!50}{9}};
		\node [below=of A-11]{\tiny \textcolor{black!50}{10}};

			\node [inner sep=1pt,above=of A-6.north east] {\footnotesize Original Data ($\file$)};


		{
			[start chain = B],
			\node (sResult1) [X,right, on chain = B] at (1.5,0) {\tiny \textcolor{blue!30}{5}};
			\node[X,right, on chain = B] {\tiny \textcolor{blue!60}{8}};
			\node(seed1)[X,right, on chain = B] {\tiny \textcolor{blue!80}{2}};
		}
		\node (PRNG1) [X, left=0.5cm of B-1, label={\tiny PRNG}, fill=black!30] {};
		\node (s1) [left=1.2cm of B-1]{$s_{1}$};

		\node [below=of B-1]{\textcolor{black!50}{\tiny $P1$}};
		\node [below=of B-2]{\textcolor{black!50}{\tiny $P2$}};
		\node [below=of B-3]{\textcolor{black!50}{\tiny $P3$}};

			\draw [arrow] (s1)  -- (PRNG1);
			\draw [arrow] (PRNG1) -- (sResult1);

		{
		[start chain = C],
		\node (sResult2) [X,right, on chain = C] at (1.5,-3.5) {\tiny 6};
		\node[X,right, on chain = C] {\tiny 1};
		\node (seed2) [X,right, on chain = C] {\tiny 3};
		}

		\node (PRNG2) [X, left=0.5cm of C-1, label={\tiny PRNG}, fill=black!30] {};
		\node (s2) [left=1.2cm of C-1]{$s_{2}$};

					\draw [arrow] (s2)  -- (PRNG2);
		\draw [arrow] (PRNG2) -- (sResult2);

		\node [below=of C-1]{\textcolor{black!50}{\tiny $P1$}};
		\node [below=of C-2]{\textcolor{black!50}{\tiny $P2$}};
		\node [below=of C-3]{\textcolor{black!50}{\tiny $P3$}};

		{
		[start chain =A1],
		\node(beginA1)[X,right, on chain = A1] at (4,0) {\footnotesize 4};
		\foreach \i in {10,8,9,1,2,12}
		\node[X,right, on chain = A1] {\footnotesize \i};
		\node (endA1) [X,right, on chain = A1] {\footnotesize 15};
		}

		\foreach \x [count=\i] in {0, ...,7}
	\node [below=of A1-\i]{\textcolor{black!50}{\tiny \x}};

		\node [inner sep=1pt,above=of A1-5.north west] {\footnotesize Outsource ($\outsource_{1}$)};

		{
		[start chain =A2],
		\node(beginA2)[X,right, on chain = A2] at (4,-3.5) {\footnotesize 4};
		\foreach \i in {1,9,7,2,2,12}
		\node[X,right, on chain = A2] {\footnotesize \i};
		\node (endA2)[X,right, on chain = A2] {\footnotesize 15};
		}

		\foreach \x [count=\i] in {0, ...,7}
		\node [below=of A2-\i]{\textcolor{black!50}{\tiny \x}};
	\node [inner sep=1pt,above=of A2-5.north west] {\footnotesize Outsource ($\outsource_{2}$)};

		{
			[start chain =A1D],
			\node(beginA1D)[X,right, on chain = A1D] at (8,0) {\footnotesize 0};
	\node[X,right, on chain = A1D] {\small $s_{1}$};
	\node[X,right, on chain = A1D] {\footnotesize \textcolor{red!30}{7}};
	\node[X,right, on chain = A1D] {\footnotesize \textcolor{red!60}{2}};
	\node[X,right, on chain = A1D] {\footnotesize \textcolor{red!80}{1}};
		}

	\node [below= of A1D-1]{\textcolor{black!50}{\tiny \invertbit}};
	\node [below=of A1D-3]{\textcolor{black!50}{\tiny  $V1$}};
	\node [below=of A1D-4]{\textcolor{black!50}{\tiny  $V2$}};
	\node [below=of A1D-5]{\textcolor{black!50}{\tiny  $V3$}};

	\node [inner sep=1pt,above=of A1D-3.north] {\footnotesize Deviation ($local_{1}$)};

		{
		[start chain =A2D],
				\node(beginA2D)[X,right, on chain = A2D] at (8,-3.5) {\footnotesize 0};
\node[X,right, on chain = A2D] {\small $s_{2}$};
\foreach \i in {1,10,8}
\node[X,right, on chain = A2D] {\footnotesize \i};
}

	\node [below= of A2D-1]{\textcolor{black!50}{\tiny \invertbit}};
	\node [below=of A2D-3]{\textcolor{black!50}{\tiny $V1$}};
\node [below=of A2D-4]{\textcolor{black!50}{\tiny $V2$}};
\node [below=of A2D-5]{\textcolor{black!50}{\tiny $V3$}};

	\node [inner sep=1pt,above=of A2D-3.north] {\footnotesize Deviation ($local_{2}$)};


	{
		[start chain = IA1]
		\node(beginIA1)[X, right, on chain= IA1] at (4,-1.7) {\footnotesize 11};
		\foreach \i in {5,7,6,14,13,3,0}
		\node[X,right, on chain = IA1] {\footnotesize \i};
	}

		\foreach \x [count=\i] in {0, ...,7}
\node [below=of IA1-\i]{\textcolor{black!50}{\tiny \x}};
	\node [inner sep=1pt,above=of IA1-5.north west] {\footnotesize Outsource ($\outsource_{3}$)};

		{
		[start chain = IA2]
		\node(beginIA2)[X, right, on chain= IA2] at (4,-5.3) {\footnotesize 11};
		\foreach \i in {14,6,8,13,13,3,0}
		\node[X,right, on chain = IA2] {\footnotesize \i};
	}

		\foreach \x [count=\i] in {0, ...,7}
\node [below=of IA2-\i]{\textcolor{black!50}{\tiny \x}};

	\node [inner sep=1pt,above=of IA2-5.north west] {\footnotesize Outsource ($\outsource_{4}$)};


		{
		[start chain =IA1D],
		\node (beginIA1D)[X,right, on chain = IA1D] at (8,-1.7) {\footnotesize 1};
		\node[X,right, on chain = IA1D] {\small $s_{1}$};
		\node[X,right, on chain = IA1D] {\footnotesize \textcolor{red!30}{7}};
		\node[X,right, on chain = IA1D] {\footnotesize \textcolor{red!60}{2}};
		\node[X,right, on chain = IA1D] {\footnotesize \textcolor{red!80}{1}};
	}

		\node [below =of IA1D-1]{\textcolor{black!50}{\tiny \invertbit}};
	\node [below=of IA1D-3]{\textcolor{black!50}{\tiny $V1$}};
	\node [below=of IA1D-4]{\textcolor{black!50}{\tiny $V2$}};
	\node [below=of IA1D-5]{\textcolor{black!50}{\tiny $V3$}};

	\node [inner sep=1pt,above=of IA1D-3.north] {\footnotesize Deviation ($local_{3}$)};

	{
		[start chain =IA2D],
		\node(beginIA2D)[X,right, on chain = IA2D] at (8,-5.3) {\footnotesize 1};
		\node[X,right, on chain = IA2D] {\footnotesize $s_{2}$};
		\foreach \i in {1,10,8}
		\node[X,right, on chain = IA2D] {\footnotesize \i};
	}

		\node [below =of IA2D-1]{\textcolor{black!50}{\tiny \invertbit}};
	\node [below=of IA2D-3]{\textcolor{black!50}{\tiny $V1$}};
	\node [below=of IA2D-4]{\textcolor{black!50}{\tiny $V2$}};
	\node [below=of IA2D-5]{\textcolor{black!50}{\tiny $V3$}};

	\node [inner sep=1pt,above=of IA2D-3.north] {\footnotesize Deviation ($local_{4}$)};

	\path [->] (endA) edge (beginA1);
	\node [rotate=42] at (3.2,-.77) {\tiny Deletions};
	\path [->] (seed1) edge (beginA1);
	\path [->] (endA) edge (beginA2);
	\node [rotate=318] at (3.2,-2.67) {\tiny Deletions};
	\path [->] (seed2) edge (beginA2);

	\path[->] ([yshift=3pt]beginA1.west) edge [bend right] (beginIA1.west);
	\node [rotate=270] at (3.95,-.77) {\tiny invert};
	\path[->] ([yshift=3pt]beginA2.west) edge [bend right] (beginIA2.west);
	\node [rotate=270] at (3.95,-4.47) {\tiny invert};


	\node (Dname) at (-1.7,-9.35) {\footnotesize $\clouddistribution$};
	{
		[start chain = PD]
		\node (Dnode1)[Y,right, on chain = PD] at(-1,-9.35) {\footnotesize 3/16};
		\node (Dnode2)[Y,right, on chain = PD] {\footnotesize 1/16};
		\foreach \i in {1/16,1/24,1/16,1/24}
		\node[Y,right, on chain = PD] {\footnotesize \i};
		\node (Dnode6)[Y,right, on chain = PD] {\footnotesize 1/24};
		\foreach \i in {3/64,1/16,1/24,1/24,3/64,1/24,3/64,3/64}
		\node[Y,right, on chain = PD] {\footnotesize \i};
		\node(endPD) [Y, right, on chain=PD] {\footnotesize 1/8};
	}
	\foreach \x [count=\i] in {0, ...,15}{
		\node [below=of PD-\i]{\textcolor{black!50}{\tiny \x}};};
	\node [below=of Dname] {\textcolor{black!50}{\tiny $symbol$}};


%
%
%
%
%

	\node (Dname2) at (-1.7,-7) {\footnotesize $\distribution_{\outsource_{4}}$};
			{
		[start chain = DA4]
		\node (DA4node1) [Y,right, on chain = DA4] at(-1,-7) {\footnotesize 1/8};
		\node (DA4node2) [Y,right, on chain = DA4] {\footnotesize 0};
		\foreach \i in {0,1/8,0,0}
		\node[Y,right, on chain = DA4] {\footnotesize \i};
		\node (DA4node6) [Y,right, on chain = DA4] {\footnotesize 1/8};
		\foreach \i in {0,1/8,0,0,1/8,0,2/8,1/8}
		\node[Y,right, on chain = DA4] {\footnotesize \i};
		\node(endDA4) [Y, right, on chain=DA4] {\footnotesize 0};
	}
		\foreach \x [count=\i] in {0, ...,15}{
	\node [below=of DA4-\i]{\textcolor{black!50}{\tiny \x}};};

	\node [below=of Dname2] {\textcolor{black!50}{\tiny $symbol$}};

	\path[->] ([yshift=-3pt]beginIA2.west) edge [bend right] ([xshift=3pt]DA4node1.north);


	\node (Cname) at (-2.5,-8.17) {\footnotesize $(\distribution_{\outsource_{4}}[i] - \clouddistribution[i])^{2}  $};

	\node (calc1) at(-0.7,-8.17) {\footnotesize 0.0040};
	\node at(-.1,-8.17) {\footnotesize +};
	\node (calc2)  at (0.6, -8.17) {\footnotesize 0.0040};
	\node at(1.2,-8.17) {\footnotesize +};
	\node at(1.6,-8.17) {\footnotesize $\cdots$};
	\node at(1.9,-8.17) {\footnotesize +};
	\node (calc3)at (2.6, -8.17) {\footnotesize 0.0070};
	\node at(3.15,-8.22) {\footnotesize =};
	\node (calcres)  at (3.7, -8.17) {\footnotesize 0.1117};

	\node (DDA1) at (11.7,0) {\textcolor{teal}{$\distance_{\outsource_{1}} = 0.2926$}};
\node (DDA2) at (11.7,-1.7) {$\distance_{\outsource_{2}} = 0.2948$};
\node (DDA3) at (11.7,-3.5) {$\distance_{\outsource_{3}} = 0.3399$};
\node (DDA4) at (11.7,-5.3) {$\distance_{\outsource_{4}} = 0.3341$};

	\path[->] (Dnode1.north) edge (calc1.south);
	\path[->] (Dnode2.north) edge (calc2.south);
	\path[->] (Dnode6.north) edge (calc3.south);
	\path[->] ([xshift=4pt]DA4node1.south) edge (calc1.north);
	\path[->] (DA4node2.south) edge (calc2.north);
	\path[->] ([xshift=-3pt]DA4node6.south) edge (calc3.north);
	\draw[->, to path={-| (\tikztotarget)}] (calcres.east) to (DDA4.south);
	\node at (8.3, -7.95) {\footnotesize Square root};



	\draw [thick] (-2.2,0.8) rectangle (2.8,-4);
	\node at (0.3, 1){\footnotesize Step1: Generating random seeds};

	\draw[thick] (3.65,0.8) rectangle (10.4,-6);
	\node at (6.97, 1){\footnotesize Step2: Generating potential outsources};

		\draw[thick] (-3.65,-6.5) rectangle (11,-10);
	\node at (3.8, -6.3){\footnotesize Step3: comparing generated outsources with Policy};

			\draw[thick] (10.6,0.8) rectangle (13,-6);
	\node [rotate= 270] at (13.2, -2.6){\footnotesize Step4: Selecting the optimal outsource};

	\end{tikzpicture}
\caption{Summary of Operations in the \user when uploading a file \file to the \cloud.}
\label{fig:client}
\end{figure*}
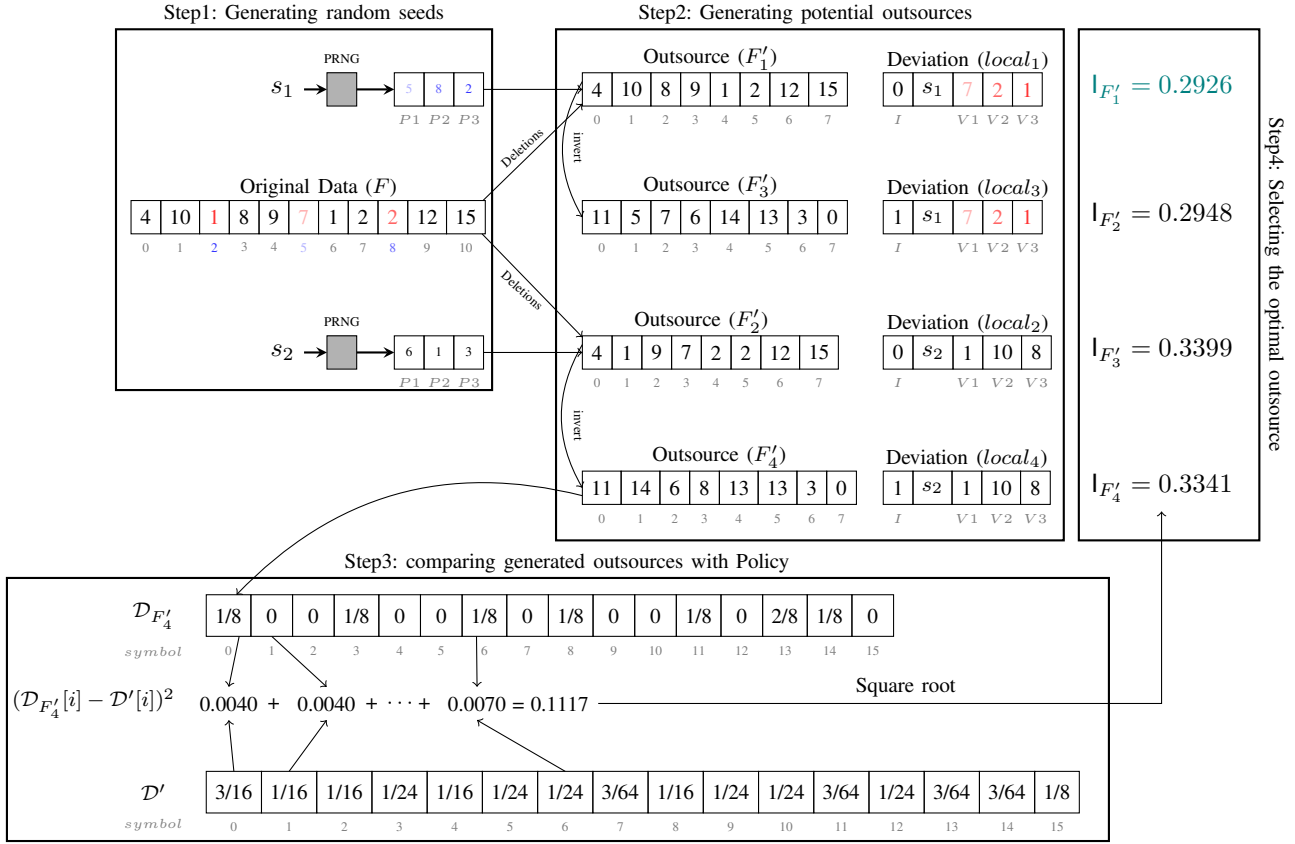

The \user that wishes to upload its data \file to the \cloud, receives a policy $\policy=[\clouddistribution, \sbase]$ that is generated by the \cloud based on the already stored data.
Following the \policy,  
the \user performs \deletions until the string \outsource has size of \sbase.
In \name, deletions are carried out on locations identified by the output of
a seeded Pseudo-Random Number Generator (PRNG).
More specifically,
the \user does not need to store the positions of the deleted symbols.
The seed of the PRNG and the deleted values suffice to reconstruct the original file \file from \outsource.
To make sure the output of \transformation fits the \policy in the best possible way,
for each file \file, the \user generates a set of \test PRNG seeds,
where \test is a \user-wide constant,
and
unique for each \user. 
If the value of \test is large, there is a higher opportunity for the
the generated \outsource to follow the probability distribution indicated in the \policy.
However, the \user has to allocate more time and computational power to 
perform the transformation for each seed.

Fig.~\ref{fig:client} illustrates a complete example of \transformation, i.e., the transformations the \user performs on \file prior to \upload.
In this example, we consider the same \file as in Fig.~\ref{fig:Del}, $k=4$, and $\test=2$, and we
assume that the policy \policy is:
\begin{equation}
	\begin{split}
		\policy = & [[\frac{3}{16},\frac{1}{16},\frac{1}{16},\frac{1}{24},\frac{1}{16},\frac{1}{24},\frac{1}{24},\frac{3}{64}, \\
		 &\frac{1}{16},\frac{1}{24},\frac{1}{24},\frac{3}{64},\frac{1}{24},\frac{3}{64},\frac{3}{64},\frac{1}{8}],\; 8]. \\
	\end{split}
	\label{eq:distribution}
\end{equation}

In this case, the \user starts by generating two random seeds, namely $\seed_{1}$ and $\seed_{2}$. 
By using $\seed_{1}$ and $\seed_{2}$ as the seeds the PRNG, the \user generates two sets of values,
each containing $\sorg-\sbase = 3$ elements.
The generated set of values indicate the position of the symbols to be deleted 
for the construction of the respective outsource strings $\outsource_{1}$ and $\outsource_{2}$.
This step is shown as step~1 in Fig.~\ref{fig:client}.
In this particular example,
$\seed_{1}$ generates the set of positions $\{5,8,2\}$ and $\seed_{2}$ generates the set $\{6,1,3\}$.

In step~2 of Fig.~\ref{fig:client}, the \user proceeds to delete the symbols in the given positions,
 storing the values of the deleted symbols in the order of deletion in $\local$. 
The result is a potential outsource $\outsource$ and a client-side deviation $\local$.
At this point,
$\local$ contains the random seed that was used to generated the \outsource and deleted values in the order that the PRNG outputs the positions.
In this particular example, considering seed $s_{1}$ generates 5, 8, and 2 in that order, the first values stored in $\local$
is the value in position 5, i.e., 7, followed by the value in position 8, which is 2 and finally the value in position 2, i.e., 1.
In order to increase the possibility of generating a string that best adheres to the probability distribution \clouddistribution in \policy, we make use of another well-known information-theoretic transformation: invert (\invert).
In the binary case, when a string is transformed by \invert, all bits that has the value of 1
 are flipped to 0 and vice-versa. More generally, \invert replaces each symbol $i$ with its complement $\alphabetSize-i-1$.
In step~2 of Fig.~\ref{fig:client} we show how the inverted outsource files are generated for a given file. 
As this example suggests, after generating the possible outsource files using \deletions,
 the \user proceeds to create the inverted counterpart of each outsource.
 In Fig.~\ref{fig:client}, $\outsource_{3}$ is generated by performing \invert on $\outsource_{1}$ and $\outsource_{4}$ is the result of \invert on $\outsource_{2}$.
 Note that as the client-side deviation is only stored locally and 
 it is not necessary for \local to follow the probability distribution of the \policy.
  Therefore, \local does not get inverted.
In order to identify if an outsource \outsource is inverted or not, the \user adds a bit to its client-side deviation \local,
denoted by \invertbit.
The value of \invertbit is equal to 1 if the \outsource has been inverted and zero otherwise.
After these actions, which conclude Step 2, the \user has a total of $2\cdot \test$ possible \outsource that can be used as the outsourced data 
to the \cloud.

The \user chooses the \outsource that best adheres the probability distribution \clouddistribution mentioned in the \policy.
In other words, the frequency of the symbols in the chosen \outsource must be the closest one to the distribution of symbols in the policy, i.e., \clouddistribution.
We define the frequency of the symbols in \outsource as

\begin{description}
	\item[Frequency of the symbols in \outsource:]
	\begin{equation}\label{eq:freq}
		\distribution^{\outsource} = [p^{\outsource}_{0}, ..., p^{\outsource}_{N-1}],
	\end{equation}
\end{description}

\begin{figure*}[!t]
	\centering
	\begin{tikzpicture}
		\node (symbolprobs) at (-2,0) {\footnotesize \begin{tabular}{|c|c|}
				\hline
				Probability & Symbol \\
				\hline
				3/16 & 0000\\
				  1/8 & 1111\\
				\vdots &\vdots \\
			3/64 & 1101 \\
				\vdots & \vdots\\
				 1/24 & 0011\\
				\hline
		\end{tabular}};

	\node () at (-2,2.4) {\tiny \begin{tabular}{c} Symbols sorted in descending\\ order of probability \end{tabular}};

		\node (table) at (2,0) {\footnotesize \begin{tabular}{|c|c|c|c|}
				\hline
				\multicolumn{4}{|c|}{\Bid} \\
				 00 & 01 & 10  & 11\\
				\hline
				0000 & 1111 & 0001 & 0010 \\
				0100 & 1000 & 1110 & 1011 \\
				1101 & 0111 & 0101 & 0110  \\
				1001 & 1010 & 1100 & 0011 \\
				\hline
		\end{tabular}};

		\draw [->] ([yshift=26pt,xshift=-13pt] symbolprobs.east) to node [above] {} ([xshift=9pt, yshift=7pt]table.west);
		\draw [->] ([yshift=-12pt,xshift=-14pt] symbolprobs.east) to node [above] {} ([xshift=9pt, yshift=-7pt]table.west);
		\draw [->] ([yshift=-42pt,xshift=-13pt] symbolprobs.east)  [bend right=10] to node [above] {} ([xshift=94pt, yshift=-31pt]table.west);

		\node (row1)[right=of table, xshift = -25pt, yshift=6pt] {\footnotesize 3/16+1/8+1/16+1/16 = \;\, 7/16};
		\node (row2)[right=of table, xshift = -25pt, yshift=-6.8pt] {\footnotesize 1/16+1/16+3/64+3/64=\quad 7/32};
		\node (row3)[right=of table, xshift = -25pt, yshift=-18.6pt] {\footnotesize 3/64+3/64+1/24+1/24 = 34/192};
		\node (row4)[right=of table, xshift = -25pt, yshift=-31pt] {\footnotesize 1/24+1/24+1/24+1/24=\;\;\; 1/6};

		\node at (6,0.7) {\tiny Sum of probabilities of each row};

		\draw [->] ([yshift=6pt,xshift=-10pt] table.east) to node [above] {} (row1.west);
		\draw [->] ([yshift=-6.8pt,xshift=-10pt] table.east) to node [above] {} (row2.west);
		\draw [->] ([yshift=-18.6pt,xshift=-10pt] table.east) to node [above] {} (row3.west);
		\draw [->] ([yshift=-31pt,xshift=-10pt] table.east) to node [above] {} (row4.west);

		\node (prob1) at (9,-0.85) {\footnotesize 11/32};
		\node (prob2) at (10,-0.5) {\footnotesize 9/16};
		\node (prob3) at (11, -0.15) {\footnotesize 1};

		\draw  ([xshift=5pt] row4.east) -- node [below] {\tiny 1} ([yshift=-1pt] prob1.west);
		\draw  ([xshift=-1pt] row3.east) -- node [above] {\tiny 0} ([yshift=1pt] prob1.west);
		\draw ([xshift=-3pt] prob1.east) -- node [below, xshift=2pt, yshift=2pt] {\tiny 1} ([yshift=-3pt]prob2.west);
		\draw  ([xshift=4pt] row2.east) -- node [above] {\tiny 0} ( prob2.west);
		\draw ([xshift=-3pt] prob2.east) -- node [below, xshift=2pt, yshift=2pt] {\tiny 1} ([yshift=-2pt]prob3.west);
		\draw  ([xshift=4pt] row1.east) -- node [above] {\tiny 0} ([yshift=1pt] prob3.west);

			\node at (9.5,0.7) {\tiny Create Huffman Coding};

		\node (table) at (12,0) {\footnotesize \begin{tabular}{|c|}
		\hline
		\Sid \\
		\\
		\hline
		0 \\
		10 \\
		110 \\
		111 \\
		\hline
\end{tabular}};

	\end{tikzpicture}

	\caption{An illustration of generating Brackets using Huffman codes for $\bits = 4$ ($\alphabet=
		\{0,1\}^4$) and $\distribution$ as in Eq.~\ref{eq:distribution}.}
	\label{fig:Huffman}
\end{figure*}
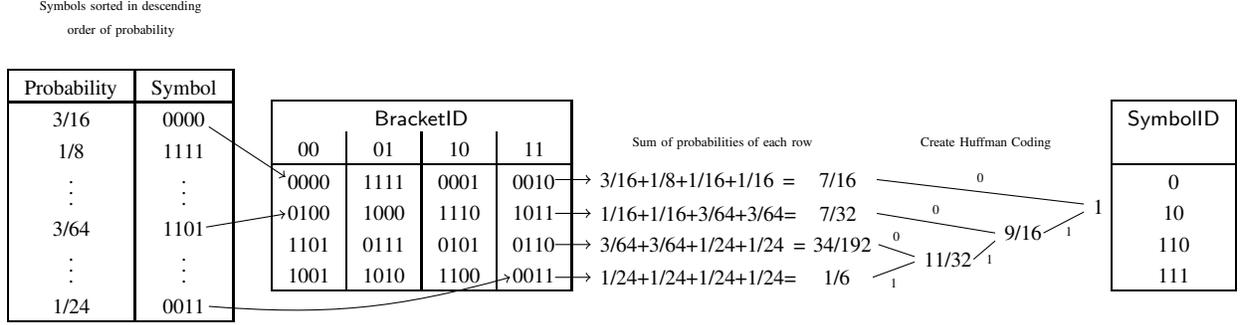
where $p^{\outsource}_{i} = \frac{\#i}{\sbase}$ for each symbol $i$, and
$\#i$ is the number of occurrences of each symbol $i$ in the \outsource.
Multiple metrics can be used in order to calculate the closeness between the frequency of symbols in a given \outsource, ($\distribution_{\outsource}$) and \clouddistribution,
e.g., Hamming distance, Euclidean distance, or Manhattan distance.
In this work, the notion of closeness between the probability distribution \clouddistribution and $\distribution_{\outsource}$
is calculated by the Euclidean distance between the two vectors,
i.e., the selected \outsource is the one with minimum value of
$\distance_{\outsource} = \sqrt{\sum\limits_{i=1}^{N} (p^{\outsource}_{i} - p^{\prime}_{i})^{2}}.$
Hence, 
as shown in Step~3 in Fig.~\ref{fig:client}, the \user first calculates the frequency of symbols in each \outsource,
creating $\distribution_{\outsource}$.
Fig.~\ref{fig:client} shows the calculated $\distribution_{\outsource_{4}}$ for $\outsource_{4}$.
We see that $\outsource_{4}$ has one element 0, therefore, $\#0 = 1$ and $p^{\outsource}_{i} = 1/8$.
After generating the distribution of symbols for all generated \outsource, the \user calculates the value of
$\distance_{\outsource}$.
This calculation is carried out for $\outsource_{4}$ in Fig.~\ref{fig:client}.
After calculating the value of $\distance_{\outsource}$ for all generated \outsource is Step 2,
the \user chooses the \outsource with the minimum value of $\distance_{\outsource}$
as the last step (Step 4).
In the given example, $\outsource_{1}$ has the lowest value of $\distance_{\outsource}$, and therefore is chosen by the \user
to be outsourced to the cloud.
Then, the client proceeds with generating \id for the selected \outsource, sending \id and \outsource to the \cloud,
while storing \id and \local locally.

\subsection{The \cloud of \name}
\label{sec:cloud}
As mentioned, the \cloud uses an innovative method 
to store the received \outsource and find possible deduplications in the received data.
In this section, we explore the ideas and theory that lead to \cloudtransformation,
and describe \cloudtransformation and deduplication in the \cloud in more detail.
We also show a toy example to illustrate
\cloudtransformation visually.

\paragraph{Motivation}
The main idea behind \cloudtransformation is to reduce the time needed to find potential deduplication between received data in the \cloud.
The idea of \cloudtransformation consists of four main ideas
 to increase the performance and compression rate in the \cloud.
These ideas include splitting data in brackets, sorting, Huffman coding and changing values. 

\begin{table}[!b]
	\caption{An illustration of brackets for $\bits = 4$, i.e., $\alphabet=
		\{0,1\}^4$. }
	\vspace{-.8em}
	\centering
	\begin{tabular}{|cc|c|c|c|c|}
		\hline
		& & \multicolumn{4}{c|}{\Bid} \\
		&  & 00 & 01 & 10 & 11 \\
		\hline
		\multirow{4}{*}{\Sid} & 00 & 0000 & 0100 & 1000 & 1100\\
		& 01 & 0001 & 0101 & 1001 & 1101\\
		& 10 & 0010 & 0110 & 1010 & 1110 \\
		& 11 & 0011 & 0111 & 1011 & 1111 \\
		\hline
	\end{tabular}
	\label{tab:brackets}
\end{table}

\textbf{Data Split using Brackets}.
This process organizes symbols in brackets. Each bracket is assigned a \Bid and each symbol in the bracket is identified with a \Sid.
The brackets are pre-defined and unique for each value of \bits.
Table~\ref{tab:brackets} gives an example of the bracket system where $\bits = 4$, where each column represents the \Bid and each row represents the \Sid of a the given symbol.
Using this table, the \cloud transforms the received outsource \outsource into two different strings:
1. A base \base which includes the \Bid of the symbols, sorted in ascending order, which is used for deduplication; and
2. An addendum \Addendum which includes the \Sid of the symbols, which is added to the cloud-side deviation.
In order to reduce the number of bracket IDs used 
we divide the aforementioned table into four zones.
Each zone is a sub-table of size $2{^\frac{k}{4}}$ . 
At its core, this approach is a sub-categorization where rarely used \Bid values are omitted and 
the symbols are fitted into fewer Brackets.
This is detailed in the Change Values section.

\textbf{Sorting}.
After performing the data split using brackets, the generated \base is sorted. 
A String \Change indicating the set of swaps performed to sort the \base is stored in cloud-side deviation.
The \base is stored in a forest data structure to ease the search process and further compress the bases.
In order to reverse the process of sorting, we store the required swaps (\swap) needed to sort \base.
Due to the split in brackets, the \cloud can perform swaps for $2{^\frac{k}{2}}$ values instead of $N=2^{k}$ values, reducing the  overhead and time complexity to store and find the required swaps.
%
%
%


\textbf{Huffman Coding.}
We leverage the fact that the \outsource received by the \cloud follows a certain probability distribution (ensured by the policy),
and use Huffman coding in order to reduce the expected size of the \Sid that the \cloud needs to store
in the cloud-side deviation.
To generate the \Sid of the symbols based on their probability distribution,
the \cloud creates a table with
 $\frac{\bits}{2}$ columns, which represent the \Bid of symbols.
Then, the \cloud populates the table by putting the symbols with highest probability distribution in
the table first.

The way that the table is populated depends on the particular structure of the data.
In this work, we populate the table by filling the rows first.
An example is illustrated in Fig. \ref{fig:Huffman}
for $\bits = 4$.
Other methods for filling the table are possible, e.g., columns first, with implications in compression and system performance.
By filling the rows first, the \cloud optimizes the size of \Sid for the symbols,
as the symbols with higher probability are assigned the same \Sid.
However, as the frequent symbols are assigned to different \Bid,
the average number of swaps would be high.
On the other hand, if the table is filled by columns first,
the expected size of \Sid is higher, but the average number of swaps is smaller.
The choice of the method and the optimal outcome for the compression rate heavily depends on the
structure and statistical characteristics of the files in \database.

After populating the table, the \cloud treats each row as a variable, and
calculates the probability of each row, which is equal to the summation of the probability of the symbols in the given row.
Then, the \cloud generates the Huffman code for the rows of the table,
and sets the value of the Huffman code as the \Sid of that row.
As an intuition, consider the same probability distribution as the example in Eq.~\eqref{eq:distribution}.
In this case, the procedure of generating the \Bid and \Sid of symbols is illustrated in Fig.~\ref{fig:Huffman}.

\begin{table}[!b]
	\caption{An illustration of brackets used for \cloudtransformation for $\bits = 8$. }
\vspace{-.8em}
\centering
\begin{tabular}{|cc|cccc|ccc|}
	\hline
	& & \multicolumn{7}{c|}{\Bid} \\
	& & 000 & 001 & $\cdots$ & 111 & 000 & $\cdots$ & 111 \\
		\hline
	\multirow{5}{*}{\Sid} & $id_{1}$ & $p_{0}$  & $p_{1}$   & $\cdots$ & $p_{7}$ & $p_{64}$ & $\cdots$ &  \\
	& $id_{2}$& $p_{8}$  & $p_{9}$  & $\cdots$ &  & $\vdots$ &  $\ddots$ &  \\
	& \vdots & \vdots &   &  $\vdots$ & $\vdots$  &  &   &  \\
	& $id_{7}$& $p_{56}$  & $p_{57}$  & $\cdots$ & $p_{63}$ & $p_{120}$ & $\cdots$ & $p_{127}$ \\
	\hline
		\multirow{3}{*}{\Sid} & $id_{1}$ & $p_{128}$  & $p_{129}$  &   $\cdots$ & $p_{135}$ & $p_{192}$ & $\cdots$ &  \\
	& \vdots&\vdots &   &  $\ddots$ & &  & $\ddots$  &  \\
	& $id_{7}$& $p_{184}$  &   & $\cdots$  & $p_{191}$ & $p_{248}$ & $\cdots$ & $p_{255}$ \\
	\hline
\end{tabular}
\label{tab:Huffman8}
\end{table}

\textbf{Change Values.} The \Sid for symbols with high probability distribution is significantly shorter than the
symbols with lower probability.
However, the size of \Sid for symbols with lower probability distribution is large, and therefore, these symbols have diminishing effect on the compression potential.
It is also beneficial to have as few \Bid as possible to reduce the number of swaps in the sorting stage.
Thus, we propose using Change Value transformation (\changev) and describe it for $\bits=8$.
Using this transformation, the \cloud changes the value of symbols with the lowest probability to symbols with the highest probability.
In order to perform this transformation, we divide the brackets table into 4 zones, as described in the Bracket stage.
Table.~\ref{tab:Huffman8} shows an example for $\bits = 8$.
In this table, we assume that the symbols are sorted based on their probability distribution,
where $p_{0}$ denotes the symbol with the highest probability, followed by $p_{1}$, and so on, e.g.,
$p_{255}$ is the symbol with the lowest probability.
The \cloud fills the table by filling the first zone, which is the zone on the top left corner,
followed by the second zone on the top right corner and then bottom left and bottom right.
In each zone, the symbols are put in their place using the rows first methods, as discussed in the Huffman coding section.
Then, the \cloud assigns a a variable for each row of all four zones,
generating the \Sid for each row as discussed earlier.
Also, each zone is assigned a Value ID \Vid.
The \Vid for each zone is generated using the same idea of \Sid.

In the end of this procedure, each symbol has a position $[i,j]$ in its zone,
an assigned \Bid, \Sid and \Vid.
Upon receiving an outsource \outsource, the \cloud first performs a change value operation.
For each symbol, the \cloud finds the position of the symbol in its zone in the table, 
Assuming the symbol is in position $[i,j]$ of one of the zones,
 the \cloud change the value of 
 that symbol
to the value in the position $[i,j]$ in the first zone (top left corner).
Then, the \cloud stores the \Vid for the symbols in a string \ChangedValues,
which is added to the cloud-side deviation.
The transformed string is stored and is used to create \base and \Addendum as discussed before.
Using this technique, the data becomes more homogenous, reducing the potential number of \Bid, and therefore reducing the expected number of swaps,
while also reducing the expected length of the \Sid for the symbols by eliminating the symbols with lower probability
distribution.
This improvement is achieved by only storing an overhead value of \Vid for each symbol,
which is smaller for symbols with higher probabiltiy distribution.

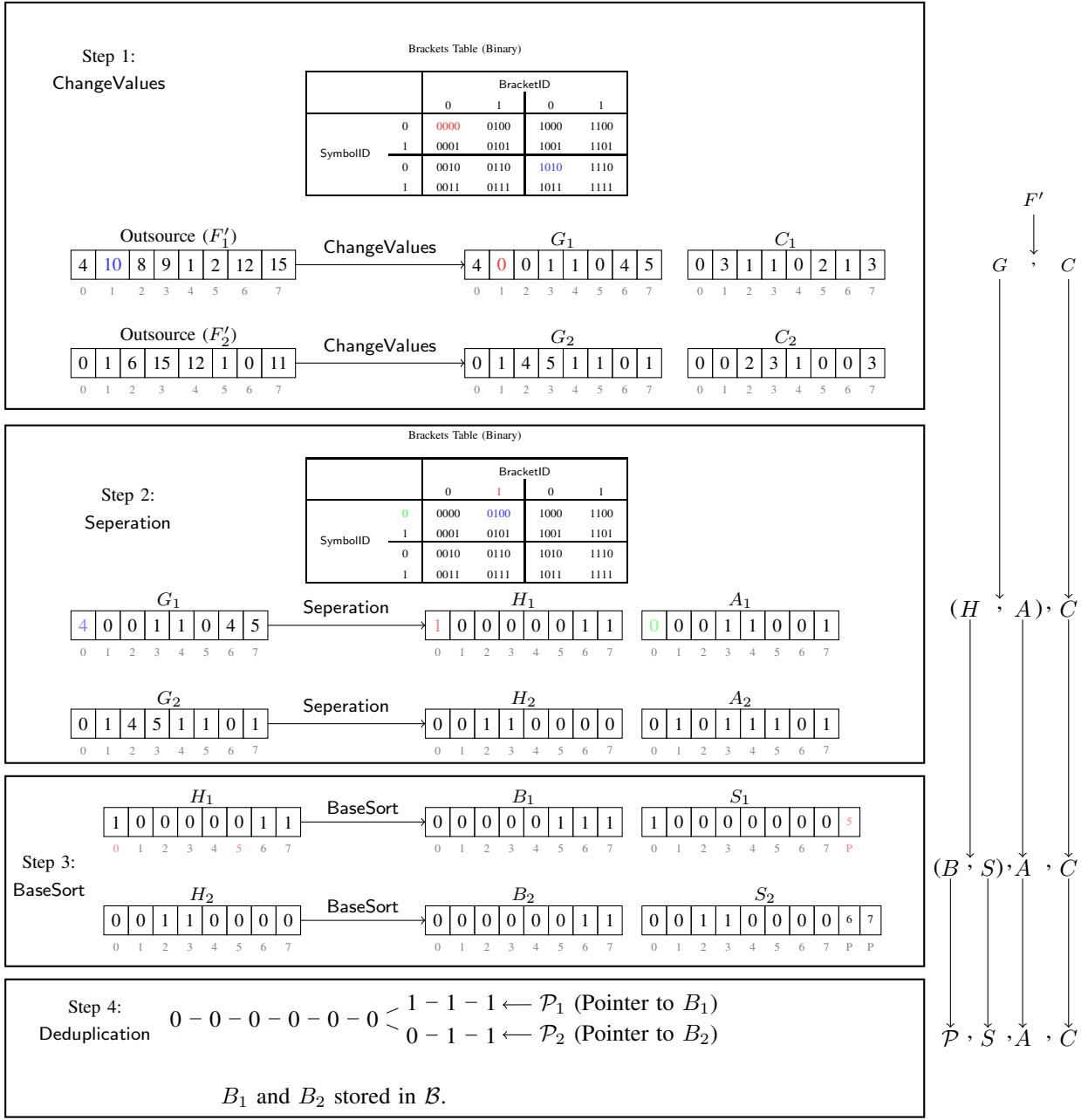
\begin{figure*}[!t]
	\centering
	\begin{tikzpicture}[
	node distance=0pt,
	X/.style = {align=left, rectangle, draw,
		minimum width=2ex, max width=2ex, minimum height=2.8ex,
		outer sep=0pt},
	Y/.style = {align=left, rectangle, draw,
		minimum width=3.5ex, minimum height=3.5ex,
		outer sep=0pt},
	]
	
	{
		[start chain =F1],
		\node(beginF1)[X,right, on chain = F1] at (-4,0) {\footnotesize 4};
		\node(exampleSymbol)[X, right, on chain = F1] {\footnotesize \textcolor{blue!80}{10}};
		\foreach \i in {8,9,1,2,12}
		\node[X,right, on chain = F1] {\footnotesize \i};
		\node (endF1) [X,right, on chain = F1] {\footnotesize 15};
	}
	
	\node [below=of F1-1]{\tiny \textcolor{black!50}{0}};
	\node [below=of F1-2]{\tiny \textcolor{black!50}{1}};
	\foreach \x [count=\i from 3] in {2, ...,7}
	\node [below=of F1-\i]{\tiny \textcolor{black!50}{\x}};
	
	\node [inner sep=1pt,above=of F1-5.north west] {\footnotesize Outsource ($\outsource_{1}$)};
	
	{
		[start chain =F2],
		\node(beginF2)[X,right, on chain = F2] at (-4,-1.5) {\footnotesize 0};
		\foreach \i in {1,6,15,12,1,0}
		\node[X,right, on chain = F2] {\footnotesize \i};
		\node (endF2)[X,right, on chain = F2] {\footnotesize 11};
	}
	
	\foreach \x [count=\i] in {0, ...,7}
	\node [below=of F2-\i]{\tiny \textcolor{black!50}{\x}};
	\node [inner sep=1pt,above=of F2-5.north west] {\footnotesize Outsource ($\outsource_{2}$)};

	%
	
	
	
	\node(BidTable) at (2,2) { \tiny{
			\begin{tabular}{|cc|cc|cc|}
			\hline
			& & \multicolumn{4}{c|}{\Bid} \\
			&  & 0 & 1 & 0 & 1 \\
			\hline
			\multirow{4}{*}{\Sid} & 0 & \textcolor{red!80}{0000} & 0100 & 1000 & 1100\\
			& 1 & 0001 & 0101 & 1001 & 1101\\
			\cline{2-6}
			& 0 & 0010 & 0110 & \textcolor{blue!80}{1010} & 1110 \\
			& 1 & 0011 & 0111 & 1011 & 1111 \\
			\hline
			\end{tabular}
	}};
	
	\node [above=of BidTable] {\tiny Brackets Table (Binary)};
	
	
	{
		[start chain =F11],
		\node(beginF11)[X,right, on chain = F11] at (2,0) {\footnotesize 4};
		\node(exampleBid)[X, right, on chain = F11] {\footnotesize \textcolor{red!80}{0}};
		\foreach \i in {0,1,1,0,4}
		\node[X,right, on chain = F11] {\footnotesize \i};
		\node (endF11) [X,right, on chain = F11] {\footnotesize 5};
	}
	
	\foreach \x [count=\i] in {0, ...,7}
	\node [below=of F11-\i]{\textcolor{black!50}{\tiny \x}};
	\node [inner sep=1pt,above=of F11-5.north west] {\footnotesize $\chnvResult_{1}$};
	
	\draw [->] (endF1) to node [above] {\footnotesize \changeVs} (beginF11);
	
	{
		[start chain =F21],
		\node(beginF21)[X,right, on chain = F21] at (2,-1.5) {\footnotesize 0};
		\foreach \i in {1,4,5,1,1,0}
		\node[X,right, on chain = F21] {\footnotesize \i};
		\node (endF21) [X,right, on chain = F21] {\footnotesize 1};
	}
	
	\foreach \x [count=\i] in {0, ...,7}
	\node [below=of F21-\i]{\textcolor{black!50}{\tiny \x}};
	\node [inner sep=1pt,above=of F21-5.north west] {\footnotesize $\chnvResult_{2}$};
	
	\draw [->] (endF2) to node [above] {\footnotesize \changeVs} (beginF21);
	
	%
	%
	%
	%
	%
	
	%
	
	{
		[start chain =C1],
		\node(beginC1)[X,right, on chain = C1] at (5.4,0) {\footnotesize 0};
		\node[X,right, on chain = C1] {\footnotesize 3};
		\foreach \i in {1,1,0,2,1}
		\node[X,right, on chain = C1] {\footnotesize \i};
		\node (endC1) [X,right, on chain = C1] {\footnotesize 3};
	}
	
	
	\foreach \x [count=\i] in {0, ...,7}
	\node [below=of C1-\i]{\textcolor{black!50}{\tiny \x}};
	
	\node [inner sep=1pt,above=of C1-5.north west] {\footnotesize $\ChangedValues_{1}$};
	
	{
		[start chain =C2],
		\node(beginC2)[X,right, on chain = C2] at (5.4,-1.5) {\footnotesize 0};
		\foreach \i in {0,2,3,1,0,0}
		\node[X,right, on chain = C2] {\footnotesize \i};
		\node (endC1) [X,right, on chain = C2] {\footnotesize 3};
	}
	%
	
	\foreach \x [count=\i] in {0, ...,7}
	\node [below=of C2-\i]{\textcolor{black!50}{\tiny \x}};
	\node [inner sep=1pt,above=of C2-5.north west] {\footnotesize $\ChangedValues_{2}$};

	\node (Line1S) at (10.67,1){\footnotesize $\outsource$};
	\node  (line1G)at (10.15,0){\footnotesize$\chnvResult$};
	\node (Line1E)at (10.67,0){\footnotesize ,};
	\node (line1CV)at (11.2,0){\footnotesize $\ChangedValues$};
	\node [text width=2cm]at (-3.5, 3) {\footnotesize \begin{tabular}{c} Step 1: \\ \changeVs \end{tabular}};
	
	\draw [->] (Line1S) -- (Line1E);
	
	

	\node(BidTable2) at (2,-3.9) { \tiny{
			\begin{tabular}{|cc|cc|cc|}
			\hline
			& & \multicolumn{4}{c|}{\Bid} \\
			&  & 0 & \textcolor{red!80}{1} & 0 & 1 \\
			\hline
			\multirow{4}{*}{\Sid} & \textcolor{green!80}{0}  & 0000 & \textcolor{blue!80}{0100} & 1000 & 1100\\
			& 1 & 0001 & 0101 & 1001 & 1101\\
			\cline{2-6}
			& 0 & 0010 & 0110 & 1010 & 1110 \\
			& 1 & 0011 & 0111 & 1011 & 1111 \\
			\hline
			\end{tabular}}
	};
	
	\node [above=of BidTable2] {\tiny Brackets Table (Binary)};
	
	
	{
		[start chain =F12],
		\node(beginF12)[X,right, on chain = F12] at (-4,-5.5) {\footnotesize \textcolor{blue!50}{4}};
		\foreach \i in {0,0,1,1,0,4}
		\node[X,right, on chain = F12] {\footnotesize \i};
		\node (endF12) [X,right, on chain = F12] {\footnotesize 5};
	}
	
	\foreach \x [count=\i] in {0, ...,7}
	\node [below=of F12-\i]{\textcolor{black!50}{\tiny \x}};
	\node [inner sep=1pt,above=of F12-5.north west] {\footnotesize $\chnvResult_{1}$};
	
	{
		[start chain =F22],
		\node(beginF22)[X,right, on chain = F22] at (-4,-7) {\footnotesize 0};
		\foreach \i in {1,4,5,1,1,0}
		\node[X,right, on chain = F22] {\footnotesize \i};
		\node (endF22) [X,right, on chain = F22] {\footnotesize 1};
	}
	
	\foreach \x [count=\i] in {0, ...,7}
	\node [below=of F22-\i]{\textcolor{black!50}{\tiny \x}};
	\node [inner sep=1pt,above=of F22-5.north west] {\footnotesize $\chnvResult_{2}$};
	%
	%
	%
	%
	
	
	{
		[start chain =H1],
		\node(beginH1)[X,right, on chain = H1] at (1.4,-5.5) {\footnotesize \textcolor{red!50}{1}};
		\foreach \i in {0,0,0,0,0,1}
		\node[X,right, on chain = H1] {\footnotesize \i};
		\node (endH1) [X,right, on chain = H1] {\footnotesize 1};
	}
	
	\foreach \x [count=\i] in {0, ...,7}
	\node [below=of H1-\i]{\textcolor{black!50}{\tiny \x}};
	\node [inner sep=1pt,above=of H1-5.north west] {\footnotesize $\sepResult_{1}$};
	\draw [->] (endF12) to node [above] {\footnotesize \sep} (beginH1);
	
	{
		[start chain =H2],
		\node(beginH2)[X,right, on chain = H2] at (1.4,-7) {\footnotesize 0};
		\foreach \i in {0,1,1,0,0,0}
		\node[X,right, on chain = H2] {\footnotesize \i};
		\node (endH2) [X,right, on chain = H2] {\footnotesize 0};
	}
	
	\foreach \x [count=\i] in {0, ...,7}
	\node [below=of H2-\i]{\textcolor{black!50}{\tiny \x}};
	\node [inner sep=1pt,above=of H2-5.north west] {\footnotesize $\sepResult_{2}$};
	\draw [->] (endF22) to node [above] {\footnotesize \sep} (beginH2);
	
	%
	%
	%
	%
	%
	
	
	{
		[start chain =A1],
		\node(beginA1)[X,right, on chain = A1] at (4.7,-5.5) {\footnotesize \textcolor{green!50}{0}};
		\foreach \i in {0,0,1,1,0,0}
		\node[X,right, on chain = A1] {\footnotesize \i};
		\node (endA1) [X,right, on chain = A1] {\footnotesize 1};
	}
	
	\foreach \x [count=\i] in {0, ...,7}
	\node [below=of A1-\i]{\textcolor{black!50}{\tiny \x}};
	\node [inner sep=1pt,above=of A1-5.north west] {\footnotesize $\Addendum_{1}$};
	
	{
		[start chain =A2],
		\node(beginA2)[X,right, on chain = A2] at (4.7,-7) {\footnotesize 0};
		\foreach \i in {1,0,1,1,1,0}
		\node[X,right, on chain = A2] {\footnotesize \i};
		\node (endA2) [X,right, on chain = A2] {\footnotesize 1};
	}
	
	\foreach \x [count=\i] in {0, ...,7}
	\node [below=of A2-\i]{\textcolor{black!50}{\tiny \x}};
	\node [inner sep=1pt,above=of A2-5.north west] {\footnotesize $\Addendum_{2}$};
	%
	%
	%
	%
	
	
	
	
	\node at (9.45, -5.25) {(};
	\node (line2H)[inner sep=1pt] at (9.7,-5.25){$\sepResult$};
	\node (line2E)[inner sep=1pt] at (10.15,-5.25){,};
	\node (line2A)[inner sep=1pt] at (10.5,-5.25){$\Addendum$};
	\node at (10.75, -5.25) {)};
	\node (line2E2)[inner sep=1pt] at (10.9,-5.25){,};
	\node (line2CV)[inner sep=1pt] at (11.2,-5.25){$\ChangedValues$};
	\draw [->] (line1G) -- ([yshift=2pt]line2E.north);
	\draw [->] (line1CV) -- (line2CV);
	
	\node [text width=2cm]at (-3, -3.7) {\footnotesize \begin{tabular}{c} Step 2: \\ \sep \end{tabular}};

	
	\node [text width=2cm]at (-4.1, -9.3) {\footnotesize \begin{tabular}{c} Step 3: \\ \baseSort \end{tabular}};
	
	{
		[start chain =H11],
		\node(beginH11)[X,right, on chain = H11] at (-3.5,-8.5) {\footnotesize 1};
		\node(swapnode1)[X,right, on chain = H11] {\footnotesize 0};
		\node[X,right, on chain = H11] {\footnotesize 0};
		\node(swapnode2)[X,right, on chain = H11] {\footnotesize 0};
		\foreach \i in {0,0,1}
		\node[X,right, on chain = H11] {\footnotesize \i};
		\node (endH11) [X,right, on chain = H11] {\footnotesize 1};
	}
	
	\node [below=of H11-1]{\textcolor{red!50}{\tiny 0}};
	\foreach \x [count=\i] in {2, ...,5}
	\node [below=of H11-\x]{\textcolor{black!50}{\tiny \i}};
	\node [below=of H11-6]{\textcolor{red!50}{\tiny 5}};
	\node [below=of H11-7]{\textcolor{black!50}{\tiny 6}};
	\node [below=of H11-8]{\textcolor{black!50}{\tiny 7}};
	\node [inner sep=1pt,above=of H11-5.north west] {\footnotesize $\sepResult_{1}$};
	
	
	{
		[start chain =H21],
		\node(beginH21)[X,right, on chain = H21] at (-3.5,-10) {\footnotesize 0};
		\foreach \i in {0,1,1,0,0,0}
		\node[X,right, on chain = H21] {\footnotesize \i};
		\node (endH21) [X,right, on chain = H21] {\footnotesize 0};
	}
	
	\foreach \x [count=\i] in {0, ...,7}
	\node [below=of H21-\i]{\textcolor{black!50}{\tiny \x}};
	\node [inner sep=1pt,above=of H21-5.north west] {\footnotesize $\sepResult_{2}$};
	
	
	{
		[start chain =B1],
		\node(beginB1)[X,right, on chain = B1] at (1.4,-8.5) {\footnotesize 0};
		\foreach \i in {0,0,0,0,1,1}
		\node[X,right, on chain = B1] {\footnotesize \i};
		\node (endB1) [X,right, on chain = B1] {\footnotesize 1};
	}
	
	\foreach \x [count=\i] in {0, ...,7}
	\node [below=of B1-\i]{\textcolor{black!50}{\tiny \x}};
	\node [inner sep=1pt,above=of B1-5.north west] {\footnotesize $\base_{1}$};
	
	{
		[start chain =B2],
		\node(beginB2)[X,right, on chain = B2] at (1.4,-10) {\footnotesize 0};
		\foreach \i in {0,0,0,0,0,1}
		\node[X,right, on chain = B2] {\footnotesize \i};
		\node (endB2) [X,right, on chain = B2] {\footnotesize 1};
	}
	
	\foreach \x [count=\i] in {0, ...,7}
	\node [below=of B2-\i]{\textcolor{black!50}{\tiny \x}};
	\node [inner sep=1pt,above=of B2-5.north west] {\footnotesize $\base_{2}$};
	
	\draw [->] (endH11) to node [above] {\footnotesize \baseSort} (beginB1);
	\draw [->] (endH21) to node [above] {\footnotesize \baseSort} (beginB2);

	
	{
		[start chain =S1],
		\node(beginS1)[X,right, on chain = S1] at (4.7,-8.5) {\footnotesize 1};
		\foreach \i in {0,0,0,0,0,0,0}
		\node[X,right, on chain = S1] {\footnotesize \i};
		\node (endS1) [X,right, on chain = S1] {\tiny \textcolor{red!50}{5}};
	}
	
	\foreach \x [count=\i] in {0, ...,7}
	\node [below=of S1-\i]{\textcolor{black!50}{\tiny \x}};
	\node[below =of S1-9]{\tiny \textcolor{red!50}{P}};
	%
	\node [inner sep=1pt,above=of S1-5.north west] {\footnotesize $\Change_{1}$};
	
	{
		[start chain =S2],
		\node(beginS2)[X,right, on chain = S2] at (4.7,-10) {\footnotesize 0};
		\foreach \i in {0,1,1,0,0,0,0}
		\node[X,right, on chain = S2] {\footnotesize \i};
		\node [X,right, on chain = S2] {\tiny 6};
		\node (endS2) [X,right, on chain = S2] {\tiny 7};
	}

	\foreach \x [count=\i] in {0, ...,7}
	\node [below=of S2-\i]{\textcolor{black!50}{\tiny \x}};
	\node[below =of S2-9]{\tiny \textcolor{black!50}{P}};
	\node[below =of S2-10]{\tiny \textcolor{black!50}{P}};
	\node [inner sep=1pt,above=of S2-6.north west] {\footnotesize $\Change_{2}$};
	
	\node at (9.2, -9.2) {(};
	\node (line3B)[inner sep=1pt] at (9.4,-9.2){$\base$};
	\node (line3E)[inner sep=1pt] at (9.7,-9.2){,};
	\node (line3C)[inner sep=1pt] at (9.97,-9.2){$\Change$};
	\node at (10.15,-9.2) {)};
	\node (line3E2)[inner sep=1pt] at (10.3,-9.2){,};
	\node (line3A)[inner sep=1pt] at (10.5,-9.2){$\Addendum$};
	\node (line3E3)[inner sep=1pt] at (10.9,-9.2){,};
	\node (line3CV)[inner sep=1pt] at (11.2,-9.2){$\ChangedValues$};
	\draw [->] (line2H) -- ([yshift=2pt]line3E.north);
	\draw [->] (line2A) -- (line3A);
	\draw [->] (line2CV) -- (line3CV);
	
	
	\node [text width=2cm]at (-3.7, -11.5) {\footnotesize \begin{tabular}{c} Step 4: \\ \dedup \end{tabular}};
	
	\node (forest) at (-0,-11.5) {\begin{tikzpicture}[level distance = 6 mm, sibling distance = 5mm, grow = right]
		\node {0}
		child {
			node{0}
			child {
				node{0} edge from parent
				child {
					node {0}
					child {
						node {0}
						child {
							node{0}
							child {
								node{0}
								child {
									node{1}
									child {
										node{1}
									}
								}
							}
							child {
								node{1}
								child {
									node{1}
									child {
										node{1}
									}
								}
							}
						}
					}
				}
			}
		};
		\end{tikzpicture}};
	
	\node at (0,-12.7) {$\base_{1}$ and $\base_{2}$ stored in \baseset.};
	
	\node (basepointer1) at (4.5, -11.28) {$\basepointer_{1}$ (Pointer to $\base_{1}$)};
	\node (basepointer2) at (4.5, -11.75) {$\basepointer_{2}$ (Pointer to $\base_{2}$)};
	
	\draw [->] (basepointer1) -- ([xshift=-4.3pt, yshift=6pt]forest.east);
	\draw [->] (basepointer2) -- ([xshift=-4.3pt, yshift=-7pt]forest.east);
	
	\node (line4P)[inner sep=1pt] at (9.4,-11.8){$\basepointer$};
	\node (line3E)[inner sep=1pt] at (9.7,-11.8){,};
	\node (line4C)[inner sep=1pt] at (9.97,-11.8){$\Change$};
	\node (line4E2)[inner sep=1pt] at (10.3,-11.8){,};
	\node (line4A)[inner sep=1pt] at (10.5,-11.8){$\Addendum$};
	\node (line4E3)[inner sep=1pt] at (10.9,-11.8){,};
	\node (line4CV)[inner sep=1pt] at (11.2,-11.8){$\ChangedValues$};
	\draw [->] (line3B) -- (line4P);
	\draw [->] (line3C) -- (line4C);
	\draw [->] (line3A) -- (line4A);
	\draw [->] (line3CV) -- (line4CV);

	
	\draw [thick] (-5,4) rectangle (9,-2.2);
	\draw [thick] (-5,-2.45) rectangle (9,-7.6);
	\draw [thick] (-5,-7.8) rectangle (9,-10.7);
	\draw [thick] (-5,-10.9) rectangle (9,-13);
\end{tikzpicture}
\caption{Example of the Transformations in the \cloud after receiving \outsource.}
\label{fig:cloud}
\end{figure*}

\paragraph{\cloudtransformation}
Let us describe in detail the journey on an \outsource as it reaches the the \cloud.
The \cloud has already generated the Brackets table before receiving \outsource.
We support our description using an example for $\bits =4$, where the Bracket table is given in Table.~\ref{tab:brackets2}.
Fig.~\ref{fig:cloud} illustrates the actions performed by the \cloud after receiving an outsource.
For the sake of simplicity, we assume that in our example, the \Vid for the zones are
$00$,$01$,$10$,$11$.

\begin{table}[!t]
	\caption{The brackets table for the toy example of Fig. \ref{fig:cloud}} 
\vspace{-.8em}
\centering
\begin{tabular}{|cc|cc|cc|}
	\hline
	& & \multicolumn{4}{c|}{\Bid} \\
	&  & 0 & 1 & 0 & 1 \\
	\hline
	\multirow{4}{*}{\Sid} & 0 & 0000 & 0100 & 1000 & 1100\\
	& 1 & 0001 & 0101 & 1001 & 1101\\
	\cline{2-6}
	& 0 & 0010 & 0110 & 1010 & 1110 \\
	& 1 & 0011 & 0111 & 1011 & 1111 \\
	\hline
\end{tabular}
\label{tab:brackets2}
\end{table}

In the first step, if $\bits= 8$, the \cloud changes the value (\changev) of the symbols to the respective symbol in the first
zone of the brackets table, while storing the value id \Vid of all symbols in a separate string denoted by \ChangedValues.
This step transforms the received \outsource to a new string denoted by \chnvResult,
which includes the values of the symbols after \changev (only 4 possible values).
This step is illustrated as Step 1 in Fig.~\ref{fig:cloud}.
As an example, the value of symbol ``10'' in position 1 is going to change.
Looking at the Brackets table, the binary representation of ``10'',
i.e., ``1010'' is in position $[0,0]$ of zone 4,
therefore, its value is changed to the symbol in position $[0,0]$ of zone 1,
which is ``0000'' in Binary,
i.e., ``0''.
The \cloud inserts the \Vid of zone 4, i.e.,  ``3'' into \ChangedValues.

After generating \chnvResult, the \cloud uses the Brackets table to separate each symbol into its \Bid and \Sid,
creating two strings.
\begin{enumerate*}
	\item String \sepResult which contains the \Bid of the symbols, and
	\item String \Addendum which contains the \Sid of the symbols.
\end{enumerate*}
This step is illustrated as step 2 in Fig.~\ref{fig:cloud}.
As an example, the first symbol in $\outsource_{1}$, i.e., $4$ 
has a $\Bid = 1$ and a $\Sid=0$. 
Therefore, the \cloud inserts 1 in the first position of \base,
and 0 in the first position of \Addendum.

\begin{algorithm}[!b]
	\caption{The algorithm to find swaps}\label{alg:cap}
	\begin{algorithmic}
		\STATE Initiate $swaps$ as an empty vector.
		\STATE $j\gets 0$
		\WHILE {$j < \sorg$}
		\IF {$\base[j] \neq \sepResult[j]$}
		\STATE Find minimum $k$ where $\base[k] = \sepResult[j] \neq \sepResult[k]$
		\STATE Swap $\sepResult[j]$ and $\sepResult[k]$
		\STATE Add $(j,k)$ to $swaps$
		\ENDIF
		\STATE $j\gets j+1$
		\ENDWHILE
	\end{algorithmic}
\end{algorithm}

In the next step, the cloud sorts \sepResult to generate \base.
We use Mergesort~\cite{cormen2009introduction} as a quick and scalable sorting algorithm in order to generate
\base from \sepResult.
In order to reconstruct \sepResult
and ultimately \outsource 
later in decompression phase,
the \cloud stores the necessary information 
about 
the difference between
\sepResult and \base.
This information is stored in \Change
as the required swaps to trasnform \sepResult into \base.
In order to identify the required swaps,
the \cloud compares \base with \sepResult,
using an algorithm inspired by Cycle Sort~\cite{cormen2009introduction},
Our algorithm does not sort \sepResult,
instead, it attempts to find the minimum number of \swap required to
transform \sepResult into \base.
In detail,
the algorithm finds the first position $j$,
where the symbol is different in \sepResult and \base, i.e.,
$\base[j] \neq \sepResult[j]$
This indicates that the value needs to be swapped in order to generate
\base.
Then, the algorithm finds the first possible position $k > j$ in \base where $\base[k] = \sepResult[j]$,
so that if the symbols in positions $j$ and $k$ are swapped in \sepResult,
the symbol in position $j$ will be equal to the corresponding symbol in \base.
In order to mitigate extra swaps,
the \cloud ensure that the symbol in position $k$ is not already in correct position by checking if the
symbol in position $k$ in \base is equal to the symbol in position $k$ of \sepResult.
i.e., the swap is only performed if
$\base[k] \neq \sepResult[k]$,
otherwise the \cloud searches for the next valid $k$.
This algorithm continues until all the symbols are in their sorted positions.
Algorithm~\ref{alg:cap} shows the pseudocode for the steps taken by our algorithm.
Note that our algorithm is sub-optimal (finding the minimal number of swaps 
to make two arrays identical is NP-hard 
\cite{gutin2014parameterized}).



After identifying the swaps,
The \cloud generates \Change.
\Change consists of two parts, a bitmap of size \sbase, and an array consisting of positions.
The \cloud starts by setting all values in the bitmap to zero,
and initiating an empty array. 
For each identified swap between two positions $i$ and $j$,
where $i <j$, the \cloud:
\begin{enumerate*}
	\item Sets the value of position $i$ in the bitmap to 1.
	\item Adds $j$ to the array that stores the positions.
\end{enumerate*}
Step 3 of Fig.~\ref{fig:cloud} shows the final result of this algorithm in our toy example.
In this particular example, for $\outsource_{1}$, only one swap is needed between the symbols in positions 0 and 5.
Hence, in \Change, the value in position 0 is set to ``1''.
The \cloud adds ``5'' to \Change, which uniquely identifies the performed
swap to be between positions ``0'' and ``5''.

The \cloud stores all bases in a set \baseset.
We leverage the fact of having sorted bases in the \cloud to create an efficient data structure to search for possible deduplications.
To achieve this, we use a forest structure to store \baseset.
The roots are the first elements of the stored bases and each child has the next element of its parents.
Each leaf represents a single \base.
The \cloud assigns a pointer to each leaf, which will indicate the \base associated to it.
Step 4 of Fig.~\ref{fig:cloud} gives an intuition of how this structure is generated,
given the \cloud only has the two bases of the \outsource shown in our example.
In our case, the forest structure is composed by single tree, due to the fact that both of the bases in the \cloud have the same initial element (with value $0$).
After creating \base,
 the \cloud searches for \base in \baseset.
If \base is found in the \baseset,
 the \cloud  deduplicates \base and store the pointer \basepointer to the \base along with the respective \Addendum, \Change and \ChangedValues of the \outsource.
Otherwise, the \cloud will add \base to \baseset, assign a pointer \basepointer to it and store \basepointer, \Addendum, \Change and \ChangedValues to represent \outsource.
Using this structure, it is easy to see that
the time complexity for searching for a \base in \baseset is $O(\sbase)$ time.
The addition of a new base to the tree takes $O(\sbase)$ time.
This is a significant improvement compared to Yggrdasil \cite{yggdrasil}, where the possible deduplications are found using brute-force search, taking up to $O(\sbase^{2}\cdot \nob)$ time.




 \paragraph{Decompression}
 When the \cloud receives a request from a \user that wants to retrieve its data,
 it reconstructs the \outsource from \basepointer, \Addendum, \Change and \ChangedValues.
 In this procedure, the \cloud simply reverses the steps taken to generate the four strings.

 \subsection{Protocol Breakdown}\label{sec:algo}

 We describe our protocol for privacy-aware dual deduplication in multi client settings by pointing out the algorithms done in \user, \cloud and between them.

 At initialization \cloud holds an initial set $n_b$-size strings called bases $\baseset=\{\base_1,\ldots, \base_b\}$.
 $\baseset$ may be empty at the start of the algorithm, in the case that the cloud has not received any data yet.
 However, $\baseset$ is updated over time by the outsourced data from the \users.


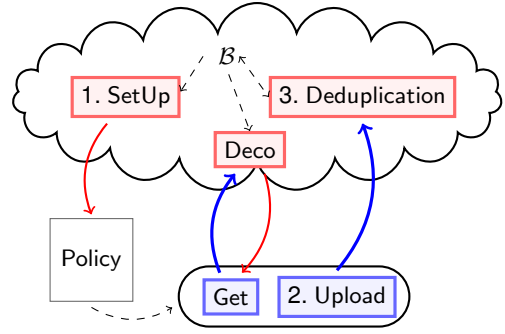
\begin{figure}[!t]
	\centering
	\begin{tikzpicture}[font=\small\sffamily, thick, scale=0.7]
	
	\node[cloud, cloud puffs=19, cloud ignores aspect, minimum width=6.5cm, minimum height=2.5cm, align=center, draw] at (0cm, 0cm) {};
	\node [alias=cloud] at (0,-1.2){};

	\node [alias=setup,style={rectangle, draw=red!60, fill=red!5, very thick}] at (-2.5,0){1. \setup};
	
	\node [alias=dedup,style={rectangle, draw=red!60, fill=red!5, very thick}] at (2,0){3. \dedup};
	
	\node [alias=set] at ([xshift=25,yshift=22] setup.east){$\baseset$};
	
	
	\node [alias=rec,style={rectangle, draw=red!60, fill=red!5, very thick}] at (-0.14,-1){\decompress};
	
	
	\draw (.7,-3.2) node[anchor=north,rounded rectangle,minimum width=3.3cm, minimum height=.8cm,
	draw, alias=one]{};
	
	\node[alias=get,style={rectangle, draw=blue!60, fill=blue!5, very thick}] at (-.5,-3.8) {{\get}};
	
	\node[alias=upload,style={rectangle, draw=blue!60, fill=blue!5, very thick}] at ([xshift=42pt]get.east) {2. \upload};
	
	\node [alias=policy, style={rectangle,  minimum width=1.1cm,minimum height=1.1cm, draw=black!60, thin}] at ([xshift=-60pt,yshift=20pt] get.west){$\policy$};
	

	\path[->, dashed,thin] ([yshift=-3pt]policy.south) edge [bend right] ([xshift=-3pt,yshift=-6pt]one.west);
	\path[->, thick,red] ([yshift=-3pt,xshift=-10pt]setup.south) edge [bend right] ([yshift=2pt]policy.north);
	
	\path[->, very thick,blue] ( [yshift=3pt]upload.north) edge [bend right] ([yshift=-3pt]dedup.south);
	
	\path[->, very thick,blue] ( [xshift=-6pt,yshift=3pt] get.north) edge [bend left] ([xshift=-8pt,yshift=-3pt]rec.south);
	
	\path[->, thick,red] ( [xshift=8pt,yshift=-3pt] rec.south) edge [bend left] ([xshift=6pt,yshift=3pt]get.north) ;
	
	\path[<->, dashed,thin] ([xshift=-3pt]set.east) edge (dedup.west);
	\path[->, dashed,thin] ([xshift=-3pt]set.west) edge ([yshift=3pt]setup.east);
	\path[->, dashed,thin] ([xshift=1pt]set.south) edge (rec.north);
	
	\end{tikzpicture}
	\vspace{-.8em}
	\caption{\name System Model for Secure, Multi-client Dual Deduplication}
	\label{fig:bonsai}
	\vspace{-1.5em}
	
\end{figure}

 \begin{description}
 	\item[$\setup(\baseset)$:]
 	This algorithm is run by the \cloud periodically.
 	It takes as input the set of bases \baseset
 	and outputs a policy $\policy = \{\sbase, \clouddistribution\}$.

 	\item[$\upload(\policy,\file)$:]
 	This algorithm is run by \user.
 	On input \policy and a file \file,
 	the \user runs local transformations \transformation
 	to generates a base \outsource of size $n_{b}$,
 	and its corresponding deviation \local.
 	Then, it generates a unique file identifier \id.
 	The algorithm outputs the triple (\id, \outsource, \local).
 	The pair (\id, \outsource) is outsourced to the \cloud,
 	while (\id, \local) is stored locally by the \user.


 	\item[$\dedup(\id, \outsource, \baseset)$:]
 	 	This algorithm is run by the \cloud.
 	 	On the input a file identifier \id, a string \outsource,
 	 	and the set of bases \baseset,
 	 	the \cloud generates the respective \base, \Addendum, \Change and \ChangedValues;
 	 the \cloud checks whether $\base\in \baseset$,
 	 in which case it stores its respective pointer, \basepointer,
 	 if not, \cloud adds \base to \baseset and assigns a pointer \basepointer to it;
 	The algorithm outputs (\id, \basepointer, \Addendum, \Change, \ChangedValues).

 	\item[$\get(\id, \local)$:]
 	This algorithm collects the \user's procedures of an interactive protocol with \cloud to retrieve an outsourced data item.
 	First, \user sends \id, symbolizing a request to retrieve the item that was outsourced with that \id.
 	Upon receiving a response \outsource from \cloud, the
 	\user uses the information encoded in \local to invert the deletions that led to \outsource, reconstructing \file.

 	\item[$\decompress(\database, \id)$:]
 	This algorithm collects the \cloud's procedures of an interactive protocol
 	to let a \user download an outsourced data item.
 	Upon receiving an \id request, \cloud checks whether \id exists in the database.
 	If not, it ignores the query.
 	Otherwise, it retrieves the corresponding record (\id, \basepointer, \Addendum, \Change, \ChangedValues);
 	inverts the deduplication performed by \dedup;
 	reconstructs the decompressed \outsource corresponding to the outsourced string; and returns \outsource back to the \user.
 \end{description}

\section{Performance Analysis}\label{sec:analysis}

In the following, we analyze the compression rate achieved by \name, the transformation cost in both \users and \cloud, and discuss the privacy achieved by \system. Unless stated otherwise, we use $\log(x)$ as the logarithm in base $2$ of $x$.

\vspace{-5pt}
\subsection{Client Compression Ratio}
\label{SS:CCR}
We begin our analysis with studying the compression ratio on the client side, i.e., \ucr in  Eq.~\eqref{eq:ucr}.
For each file \file, the user stores the respective \local and \id.
To compute \ucr for one file \file, we accurately measure the size of \local.
As the \user uses a PRNG to identify the position of the deleted elements,
the required size to store the positions is equal to the size of the PRNG seed.
we denote this size by \sseed.
The \user performs $\nodel = \sorg - \sbase$ subsequent \deletions on \file,
where the \user stores the value of all deleted symbols in \local.
Thus, \local contains a total of $\nodel\cdot \bits$ bits to store the value of the deleted symbols.
\local also contains 1 bit, indicating whether the \invert transformation has been applied to \outsource.
Therefore,
the required storage needed in \user is equal to:
\begin{equation*}
\size(\user_{i}, \file) = \sseed + \nodel\cdot \bits + \sizefid + 1,
\end{equation*}
where $\sizefid$ is the size in bits of a file identifier.

If the database for the client $i$, i.e., $\database_{i}$ has \nof files, the required storage on the \user side is:
\begin{equation*}
\size(\user_{i}, \database_{i}) = \nof\cdot (\sseed + \nodel\cdot \bits + \sizefid + 1).
\end{equation*}
The original size required for $\database_{i}$ prior to the transformations of \name is:
\begin{equation*}
|\database_{i}| = \nof\cdot \bits\cdot n_{o}.
\end{equation*}

Thus,
\begin{equation}
\ucr = \frac{\size(\user_{i}, \database_{i})}{|\database_{i}|} = \frac{\sseed + \nodel\cdot \bits + \sizefid + 1}{\bits\cdot \sorg}.
\label{eq:ygg_ucr}
\end{equation}

\subsection{Cloud Compression Ratio}
The compression rate on the \cloud, i.e., \ccr, according to Eq.~\eqref{eq:ccr}, considering the different contributing effects.
As described is Section~\ref{sec:contribution}, the cloud stores
three strings \Addendum, \Change, and \ChangedValues for each received \outsource, as well as the identifier \id and a pointer to the base \base.
The bases are stored in \baseset, where we denote the size of \baseset as the size of the forest \sizeforest used to represent all stored bases
Note that $\sizeforest < \sbase\cdot |\baseset|$, as the forest structure removes the redundancy between different bases in \baseset.
Thus, the total storage in the cloud is equal to the cost of storing \Addendum, \Change, \ChangedValues, the \id and a pointer to the respective \base for each received \outsource plus the required storage for the \baseset.
Our experiments show that \sizeforest is very small compared to the size of the data
as the \cloud is able to perform deduplication on a large number of bases.

The expected size of the \Addendum is equal to the expected length of Huffman Coding, which is equal to the entropy of the data in the \cloud, i.e., $H(\cloud)$.
The size of \Change is dependent on the number of swaps needed to sort the base \base.
For each \swap, the cloud needs to store 1 position for the position of the swapped elements,
and each position requires $\log\sbase$ space.
\Change also includes a bitmap of symbols, which has the size of \sorg bits.
Therefore, the \cloud requires $\sorg + \nswap \cdot \log\sbase$ storage for \Change,
where \nswap is the number of swaps required to sort \sepResult.
The number of swaps \nswap is heavily dependant on the structure of the data and the
way the files are chunked from the database \database.
In \ChangedValues the \cloud stores the Huffman representation of the zones, so the size of this string is
also dependent on the structure and the probability distribution of the data.
In worst case, the size of \Vid for each symbol is exactly 2 bits, leading to $2\sorg$ required storage for \ChangedValues.
In the following analysis, we use this value.
In practice, the \cloud uses Huffmann codes to determine the size of \Vid,
so the size of \Vid for most probable symbols is usually 1 bit,
leading to
lower rquired storage for use cases. 

For each \outsource, we have:
\begin{equation*}
	\begin{split}
		 E(\size(\cloud, \file)) = & H(\clouddistribution) + 3\sorg + \nswap\cdot\log\sbase +\\
		& \sizefid + \sizep,
	\end{split}
\end{equation*}
where \sizep is the size of the pointer or identifier to the \base in \baseset.

Thus, the total compression rate on the cloud is
\begin{equation}
	\footnotesize
		\ccr = \frac{\sizeforest + \nof(H(\clouddistribution) + 3\sorg + \nswap\cdot\log\sbase + \sizefid + \sizep)}{\nof\cdot\bits\cdot\sorg}.
	\label{eq:ygg_ccr2}
\end{equation}

%
%
%
\subsection{Global Compression Ratio}
The required storage size in \user side does not change and can be calculated as \size(\user, $\database_{i}$) in section~\ref{SS:CCR}.
Therefore, the global compression ratio of the system is given by the sum of the \user compression ratio and the \cloud one

\begin{equation}
	\footnotesize
		\compratio = \frac{\sizeforest + \nof(\cons + H(\clouddistribution) + \nodel\cdot \bits + 3\sorg + \nswap\cdot\log\sbase)}{\nof\cdot k\cdot \sorg},
\label{eq:C1}
\end{equation}
where \cons is a constant number, denoting the size of the pointers and the random seed of the PRNG, and is equal to $\cons = \sseed + \sizep + 2\sizefid + 1$.

\subsection{Transformation Costs}
In this section, we calculate the computation cost of the algorithms discussed in Section~\ref{sec:contribution} in \cloud and \user.
We provide two theorems for the computational cost of operations in \cloud and \user.

\begin{theorem}
The computational cost of the procedures applied in the \user during \upload is $O(\test\cdot \sorg)$.
\end{theorem}
\begin{IEEEproof}
The \user performs \deletions for each file prior to upload, which is in the \upload procedure.
The \user performs two set of instructions in \upload.
First, performing \deletions for each file and PRNG seed.
Second, calculating the probability distribution of the generated \outsource to choose the one that adheres the best with the distribution \clouddistribution provided in the \policy.

Using a vector data structure for file \file, results in linear complexity for deletion of $\sorg-\sbase$ elements in the size of the \file, i.e., $O(\sorg)$. This can be performed by copying elements not scheduled for deletion to a different vector of size $\sbase$, skipping the copy of deleted elements. Since the procedure is performed $\test$ times for \test different seeds,
 the total complexity of this instruction is $O(\test\cdot \sorg)$.


For each generated \outsource and its inverted \outsource, the \user must calculate the probability distribution. Calculating the probability distribution requires reading all the symbols, i.e., $O(\sorg)$ operations. As we have $2\cdot\test$ possible \outsource to evaluate, this instruction has compelxity $O(\test\cdot \sorg)$.
Thus, the complexity at the \user is $O(\test\cdot \sorg)$.
\end{IEEEproof}

\begin{theorem}
	The computational cost of procedure in the \cloud after receiving an \outsource form a \user during \dedup is equal to $O(\sbase(\log\sbase+\alphabetSize))$ when using MergeSort for the Sorting phase.
\end{theorem}

\begin{IEEEproof}
	The cloud performs three sets of transformations on the received \outsource, namely \changev, \sep and \sort.
	Therefore, in order to calculate the time complexity at the \cloud after receiving an \outsource, we calculate the time complexity of these transformations.

	In \changev, the \cloud needs to (1) find the symbols in the Brackets table and (2)change the value of the symbol and insert the new value, as well as \Vid in the two output strings \chnvResult and \ChangedValues.
	 The most efficient way to perform this transformation is to search for each symbol in the outsource \outsource
	 instead of searching for each symbol in the table.
	Searching for a symbol has a time complexity of $O(\sorg)$.
	As the \cloud has to search for $\alphabetSize$ symbols in total, this step has complexity $O(\sorg\cdot\alphabetSize)$.
	As we use vector data structure, the \cloud can create the two output strings in linear time, i.e., $O(\sbase)$.
	Therefore, the total time complexity of \changev is equal to $O(\sbase\cdot\alphabetSize)$.

	During \sep, the cloud creates 2 strings \Addendum and \sepResult from \chnvResult.
	For each symbol, the \cloud needs to find the respective value of \Sid and \Bid from the Brackets Table.
	This process requires a brute-force search, and therefore has a time complexity of $O(\alphabetSize)$.
	Populating both strings is done in linear time complexity after finding the \Sid and \Bid of each symbol.
	Thus, the total time complexity of this step is equal to $O(\sbase\cdot \alphabetSize)$.

	The computational cost of the transformations applied in the \cloud during \sort depends on the sorting algorithm that is used.
	In this work, we use Mergesort in order to sort the \sepResult, which has a time complexity of $O(\sbase\log\sbase)$.
	Then, the \cloud finds the swaps required using a linear comparison between the sorted base \base and \sepResult.
	As this algorithm is linear, it has a time complexity of $O(\sbase)$.
	Therefore, in total, the time complexity of \sort is equal to $O(\sbase\log\sbase)$.

	Combining these three time complexities gives the final time complexity of the \dedup procedure.
\end{IEEEproof}

%
%
%
%

\begin{theorem}
	The computational cost of the transformations applied during the client in \get is $O(\sorg)$.
\end{theorem}
\begin{IEEEproof}
	The \user needs to insert the deleted values in \local into their respective positions in the \outsource.
	This is linear in cost with the size of $\sorg$.
\end{IEEEproof}

\begin{theorem}
	The computational cost of the transformations applied during the client in \decompress is $O(\nob + \sbase\cdot\alphabetSize)$.
\end{theorem}
\begin{IEEEproof}
	The cloud performs \decompress in four steps. First, retrieving the \base from \baseset. Second, undoing the \swap transformations. Third, recreating the symbols from their respective \Bid and \Sid. Lastly, restoring the changed values.

	Retrieving the base from \baseset requires a search on the received \id.
	Although the bases are stored in a structured way, there is no guarantee that the \id is stored in the same way.
	Therefore, searching for a \id requires linear time dependent on the number of bases in the baseset, i.e., $O(\nob)$.

	After retrieving the \base, the \cloud needs to reverse the swaps stored in \Change to retrieve \sepResult by first  identifying the required swaps. This is done by a read through \Change,
	that has linear complexity, i.e., $O(\sbase)$.
	Then, the cloud performs the swaps.
	In the worst case,the number of swaps is linear to the size of the \base, i.e., $O(\sbase)$.

	Restoring the \outsource from \sepResult requires replacing each triple of \Sid, \Bid and \Vid with its respective symbol.
	Therefore, for each symbol in \sepResult, the \cloud needs to scan through the Brackets table to find the value associated
	with the triple. In the worst case scenario, the \cloud needs to scan through the whole table for each symbol, leading
	to a time complexity of $O(\alphabetSize)$ for each symbol. Therefore,
	in worst case, the time complexity of this action is equal to $O(\sbase\cdot\alphabetSize)$.

	Combining these three time complexities concludes the proof.
\end{IEEEproof}

\subsection{Privacy Analysis}\label{sec:security_proofs}
In the following, we calculate \bonsai's leakage \leakage and uncertainty \unmetric for an honest-but-curious \cloud.
In detail, we consider \Wadversary and \Sadversary as introduced in Section \ref{sec:sec_model}, and for each adversary type, we investigate two scenarios. First, the adversary has negligible probability of breaking the PRNG. Second, the adversary breaks the PRNG.

\subsubsection{Weak Adversary}
 Recall that \Wadversary has no information about the distribution of original files. This means that in its perspective, all possible values of $\file$ have the same probability of being the \user's data.
Therefore, $H(\file) = \bits\cdot \sorg$.
In scenario 1, the PRNG behaves like a truly random function in \Wadversary's view, and therefore it leaks no information about the position of the deleted elements.
Given $F^{\prime}$, the original string \file can be any of the possible strings that generate $F^{\prime}$. As \Wadversary does not have any extra information, each potential \file has the same probability of being the original string, so $H(\file|\outsource) = log(\nopreim)$, where \nopreim is the number of possible pre-images of \outsource.
 as it has been shown in~\cite{yggdrasil}, we have:
 \begin{equation*}
 	\nopreim = \sum_{j=0}^{\sorg-\sbase} {\sorg \choose j+\sbase} (2^{\bits}-1)^{\sorg-\sbase-j}.
 \end{equation*}

Therefore, the leakage for \Wadversary is
\begin{equation*}
	\leakage(F^{\prime}) = \frac{\bits\cdot \sorg - \log(\nopreim)}{\bits\cdot \sorg},
\end{equation*}
and the uncertainty \unmetric that the \Wadversary faces after receiving \outsource is
\begin{equation*}
	\unmetric = \log(\nopreim).
	\end{equation*}

In scenario 2, \Wadversary breaks the PRNG, this means that it gains knowledge about the position of the deleted symbols.
Therefore, the total number of possible pre-images of the \outsource, \nopreim has a different value.
However, the adversary still cannot distinguish between the different possible pre-images, as in the adversary's viewpoint,
all possible pre-images has the same probability of being the original data.
In order to generate a possible pre-image, the adversary can insert a symbol in the positions that the deletions has occurred.
As there are $\sorg-\sbase$ positions and each position has $2^{\bits}$ possible values as the symbol,
the number of pre-images of the \outsource is equal to:

  \begin{equation*}
 	\nopreim = 2^{\bits(\sorg-\sbase)}.
 \end{equation*}

 The rest of the analysis is the same as the previous case, therefore, if the \Wadversary breaks the PRNG, we have the following value of leakage after receiving the \outsource:
 \begin{equation*}
 	\leakage(F^{\prime}) = \frac{k\cdot \sorg - \log(m)}{k\cdot \sorg} = \frac{\sbase}{\sorg}.
 \end{equation*}

The Uncertainty for the attacker in this case is equal to:
 \begin{equation*}
 	\unmetric = \log(\nopreim) = \bits(\sorg-\sbase).
 \end{equation*}

\subsubsection{Strong Adversary}
Recall that \Sadversary has knowledge about the probability distribution of symbols in the client.
We further assume here that the adversary has a negligible probability of breaking the PRNG, i.e., has no information about the position of the deleted symbols.
 In this case, $H(\file) \neq \bits\cdot \sorg$.
 When the cloud receives \outsource, the original string \file can be any of the possible strings that generate \outsource. The cloud can generate all possible values of \file.
 The probability that a given \singlefile is the actual data of the \user is higher if 1.\singlefile has higher probability based on \distribution and 2.\singlesource can be generated by \singlefile with various deletions. We define $W(\file = \singlefile|\outsource = \singlesource)$ is the number of ways that \outsource can be generated from \file, i.e., number of distinct occurrences of \singlesource in \singlefile as a subsequence. This value is calculated using a recursive algorithm and dynamic programming.
 Using this variable, the probability of each \file in cloud's perspective is equal to:
 \begin{equation*}
 	P(\file = \singlefile|\outsource=\singlesource) = \frac{W(\file = \singlefile|\outsource = \singlesource) P(\file = \singlefile)} {\sum\limits_{\singlefile}W(\file = \singlefile|\outsource = \singlesource)P(\file = \singlefile)}.
  \end{equation*}

 Using the value of $P(\file = \singlefile|\outsource=\singlesource)$, we calculate the value for $H(\file|\outsource)$ which is equal to the uncertainty of the \Sadversary after it receives \outsource. The leakage of information to \Sadversary is calculated using the value of \unmetric and Eq.~\eqref{eq:leakage}.

If the adversary breaks the PRNG, this analysis still holds, and we need to calculate the value of $P(\file = \singlefile|\outsource=\singlesource)$ for each possible pre-image.
However, the difference between the two cases is the more information that the adversary possesses if it breaks the PRNG means that the number of possible pre-images is lower,
therefore, after receiving the \outsource, on average more information is leaked to the \cloud.

The uncertainty provided against a strong adversary
is upper bounded by the uncertainty of the weak adversary,
i.e., $k(\sorg-\sbase)$.
This fact is useful to argue for privacy against a real adversary,
which lies somewhere between the weak and strong adversaries.
In Section~\ref{sec:experiments} we calculate the values of leakage and uncertainty
against weak and strong adversaries and
discuss the implications of it in real life scenarios.

\section{Simulation Results and Discussion}\label{sec:experiments}

In this section, we show and discuss the performance of \name in terms of its compression rate and privacy.
We use three different real world datasets, including, $14$~GB of email texts in enron-mail dataset \cite{shetty2004enron}, $18$~GB of Hadoop Data File System (HDFS) logs \cite{he2020loghub}
and $10$~GB of DeVice Independant (DVI) files \cite{10.3389/fncir.2019.00005}.
We developed a C++ implementation consisting of a \name client applying deletions and a server performing deduplication algorithm described in this work.
For the sake of storage friendly implementation, we defined the \id of bases to be a global variable auto-incremented by the Cloud.
This is also used for the pointers to the bases in the \cloud.

\begin{figure}[!t]
	\centering
	\begin{tikzpicture}[
	scale=0.7,
	every axis/.style={ ybar stacked,
		ylabel = {\compratio},
		xlabel = {\bits},
		xmin={[normalized]0},
		xmax={[normalized]3},
		xtickmax={[normalized]2},
		xtick distance = 1,
		ymin = 0,
		ymax = 1,
		symbolic x coords={
			0, $4$, $8$
		},
		bar width=8pt,
		legend style={legend columns=-1,at={(0.267,0.93),},anchor=west},
	},
	]

	\begin{axis}[bar shift=-13.5pt, hide axis]
	\addplot+[fill=green] coordinates {($4$,0.7351)($8$,0.7045)};
	\addplot+[fill=blue] coordinates {($4$,0.0082)($8$,0.0082)};
	\legend{\strut \cloud, \strut \user}
	\end{axis}

	\begin{axis}[bar shift=-4.5pt, hide axis]
	\addplot+[fill=green] coordinates {($4$,0.8125)($8$,0.6391)};
	\addplot+[fill=blue] coordinates {($4$,0.0082)($8$,0.0082)};
	\legend{\strut \cloud, \strut \user}
	\end{axis}

	\begin{axis}[bar shift=0pt]
	\end{axis}

	\begin{axis}[bar shift=4.5pt, hide axis]
	\addplot+[fill=green] coordinates {($4$,0.8333)($8$,0.7616)};
	\addplot+[fill=blue] coordinates {($4$,0.0082)($8$,0.0082) };
	\end{axis}

	\begin{axis}[bar shift=13.5pt, hide axis]
	\addplot+[fill=green] coordinates {($4$,0.7904)($8$,0.7391)};
	\addplot+[fill=blue] coordinates {($4$,0.0082)($8$,0.0082) };
	\end{axis}
	\end{tikzpicture}
	\vspace{-1em}
	\caption{The compression rate in the \cloud and the \user for $\bits\cdot\sorg = 2048 bits$ and $\bits\cdot\sbase = 1928 bits$. The first column is for HDFS dataset. The second column is for DVI dataset, the third column is for enron-mail and the fourth column is when all three datasets are used.}
	\label{fig:result1}
\end{figure}
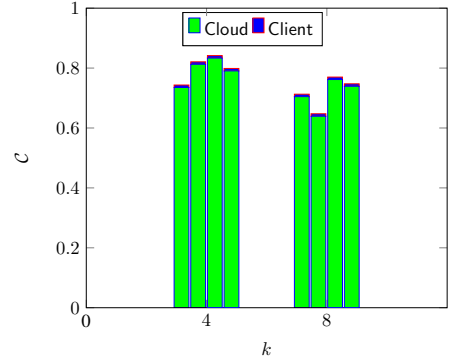

\textbf{Compression rate for different symbol size:}
Fig.~\ref{fig:result1} shows the compression rate of \bonsai for
the three datasets for two different symbol size of $\bits=4$ and $\bits=8$.
For $\bits = 8$, \name achieves a total compression rate of $\compratio = 0.6473$ for the DVI dataset,
where the \cloud achieves a compression rate of $\ccr = 0.6391$,
and the \user stores $\ucr = 0.0082$ of the data.
For the same dataset, the total compression rate
for $\bits = 4$ is equal to $\compratio = 0.8207$,
where the \cloud and the \user store
$\ccr = 0.8125$ and $\ucr = 0.0082$ of
the data respectively.
We note that \name does not drop significantly in compression potential
by using diverse data, as the total compression rate for all three datasets is
$\compratio = 0.7473$ when $\bits = 8$ and $\compratio = 0.7986$ when $\bits=4$.
This is an interesting characteristic of \name which shows the potential to outperform  other compression techniques in heterogenous datasets or workloads (e.g., mixed data sources with different characteristics and statistics).

Fig.~\ref{fig:result1} also suggests that for a fixed size of the original file $\bits\cdot\sorg = 2048$,
the total compression rate and the compression rate on the \cloud is better when the size of symbols is set as $k=8$.
This is because
(1) \name cannot reduce the size of \Sid and \Vid using Huffman coding when $\bits=4$,
as opposed to  $k=8$ leading to more storage requirement for \Addendum and \ChangedValues
when $k=4$; and,
(2) there are more symbols when $k=4$, therfore, the \cloud needs to assign more storage
for the positions when storing swaps in \Change.

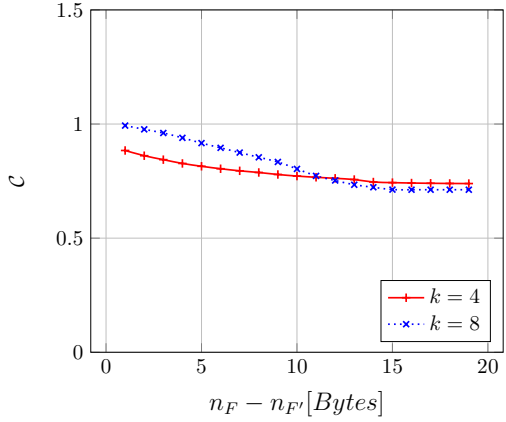
\begin{figure}[!t]
	\centering
	\begin{tikzpicture}
	[scale=0.8]
	\begin{axis}[
	xlabel= {$\sorg-\sbase [ Bytes] $},
	ylabel=$\compratio$,
	xlabel style={font=\large},
	ymin =0,
	ymax = 1.5,
	grid=major,
	legend pos= south east]

%

	\addplot[mark=+, smooth ,red, thick] plot coordinates {
		(1, 0.8841) 
(2, .8612) 
(3, .8439)
(4, .8273)
(5, .8149)
(6, .8041)
(7, .7950)
(8, .7877)
(9, .7789)
(10, .7720)
(11, .7667) 
(12, .7621) 
(13, .7568)
(14, .7459)
(15, .7433)
(16, .7415)
(17, .7404)
(18, .7396)
(19, .7392)
	};
	\addlegendentry{$k=4$}

	\addplot[mark=x, smooth, dotted, blue, thick, mark options={solid}] plot coordinates {
		(1, 0.9932) 
		(2, .9767) 
		(3, .9607)
		(4, .9395)
		(5, .9164)
		(6, .8957)
		(7, .8749)
		(8, .8542)
		(9, .8339)
		(10, .8033)
		(11, .7728) 
		(12, .7527) 
		(13, .7339)
		(14, .7230)
		(15, .7127)
		(16, .7122)
		(17, .7122)
		(18, .7122)
		(19, .7122)
	};
	\addlegendentry{$k=8$}

	\end{axis}
	\end{tikzpicture}
	\caption{Compression rates for HDFS dataset for different values of $\sbase$ when $\sorg = 256$. (Lower is better.)}
	\label{fig:results0}
\end{figure}

\textbf{Compression rate for different number of deletions:}
Fig.~\ref{fig:results0} shows the behavior of \name for different number of deletions in the \user side and its effect on the total compression rate.
As this figure suggests, increasing the number of deletions in the \user, initially results in
\name gaining in terms of total compression rate.
This is due to the fact that a larger number of deletions have two effects on the outsourced data.
First, it reduces the size of the outsourced data, allowing for more potential deduplications and lower number of required swaps.
Second, it alters the probability distribution of the outsourced data more, giving more chance to have a closer probability distribution as the distribution indicated in the policy.

However, after a certain threshold 
(14 deletions for $\bits=4$ and 15 deletions for $k=8$),
this trend does not continue and the compression rate stays the same.
This is due to the fact that after this threshold, the compression gain on the \cloud
is quite low, and is offset by higher storage requirements in the \user.
As the \cloud does not compress \Addendum and \Change, which
consist of \Sid and \Vid of the symbols, there is a lower bound on the compression rate achievable by the \cloud.
After this point, the compression rate on the \cloud improves this improvement, but it is equal to or less than the loss of compression in the \user, leading to a roughly constant total compression rate.
This behavior can be seen for both cases of $\bits = 4$ and $bits=8$.
However, in the case of $\bits=8$ the compression rate declines faster
by increasing the numjber of deletions,
compared to $\bits=4$;
as deletion of the symbols with low probability
distribution has a higher effect on reducing the number of required swaps; and
shaping the probability distribution of the symbols,
leading to more optimal values for \Sid and \Vid.

\begin{figure}[!t]
	\centering

	\subfloat[HDFS]{\begin{tikzpicture}
		[scale=0.7]
		\begin{axis}[
		xlabel= {$\bits\cdot \sorg [Bits]$},
		ylabel=$\compratio$,
		xlabel style={font=\large},
		xtick = {0,1,2,3,4,5,6,..., 16},
		xticklabels = {,,,,,$2^{5}$,$2^{6}$,$2^{7}$,$2^{8}$,$2^{9}$,$2^{10}$,$2^{11}$,$2^{12}$,$2^{13}$,$2^{14}$,$2^{15}$,$2^{16}$},
		ymin=0,
		ymax = 1.5,
		grid=major,
		legend pos= north west]

		\addplot[mark= +, smooth ,black, thick] plot coordinates {
			(5, 0.49) 
			(6, 0.63) 
			(7, 0.68)
			(8, 0.65)
			(9, 0.66)
			(10, 0.65)
			(11, 0.68)
			(12, 0.73)
			(13,0.78)
			(14,0.85)
			(15,0.89)
			(16,0.91)
		};
		\addlegendentry{$MKRE$}

		\addplot[mark = o, smooth, violet, thick] plot coordinates {
			(5, 0.60)
			(6, 0.63) 
			(7, 0.64)
			(8, 0.65)
			(9, 0.68)
			(10, 0.72)
			(11, 0.74)
			(12, 0.77)
			(13,0.80)
			(14,0.84)
			(15,0.88)
			(16,0.92)

		};
		\addlegendentry{Yggdrasil}

%
%
%
%
%
%
%

		\addplot[mark=x, smooth,blue, thick] plot coordinates {
			(5, 0.68)
			(6, 0.71) 
			(7, 0.70)
			(8, 0.69)
			(9, 0.68)
			(10, 0.72)
			(11, 0.71)
			(12, 0.69)
			(13,0.72)
			(14,0.77)
			(15,0.82)
			(16,0.85)
		};
		\addlegendentry{\name}

%
%
%
		\end{axis}
		\end{tikzpicture}}
	\\
	\subfloat[DVI]{\begin{tikzpicture}
		[scale=0.7]
		\begin{axis}[
		xlabel= {$\bits\cdot\sorg [Bits]$},
ylabel=$\compratio$,
xlabel style={font=\large},
xtick = {0,1,2,3,4,5,6,..., 16},
xticklabels = {,,,,,$2^{5}$,$2^{6}$,$2^{7}$,$2^{8}$,$2^{9}$,$2^{10}$,$2^{11}$,$2^{12}$,$2^{13}$,$2^{14}$,$2^{15}$,$2^{16}$},
ymin=0,
ymax = 1.5,
grid=major,
legend pos= north west]

		\addplot[mark= +, smooth ,black, thick] plot coordinates {
			(5, 0.42) 
			(6, 0.39) 
			(7, 0.37)
			(8, 0.35)
			(9, 0.34)
			(10, 0.33)
			(11, 0.33)
			(12, 0.33)
			(13,0.38)
			(14,0.42)
			(15,0.48)
			(16,0.52)
		};
		\addlegendentry{$MKRE$}

		\addplot[mark = o, smooth, violet, thick] plot coordinates {
(5, 0.54)
(6, 0.58)
(7, 0.62)
(8, 0.69)
(9,0.76)
(10,0.82)
(11,0.87)
(12, 0.96)
(13, 1.03)
(14, 1.09)
(15, 1.17)
(16, 1.26)

		};
		\addlegendentry{Yggdrasil}

		\addplot[mark=x, smooth,blue, thick] plot coordinates {
(5, 0.54)
(6, 0.56)
(7, 0.58)
(8, 0.64)
(9,0.69)
(10,0.74)
(11,0.78)
(12, 0.83)
(13, 0.87)
(14, 0.91)
(15,0.94)
(16,0.97)

		};
		\addlegendentry{\name}


		\end{axis}
		\end{tikzpicture}}
	\\
	\subfloat[Enron-mail]{\begin{tikzpicture}
		[scale=0.7]
		\begin{axis}[
		xlabel= {$\bits\cdot\sorg [ Bits]$},
ylabel=$\compratio$,
xlabel style={font=\large},
xtick = {0,1,2,3,4,5,6,..., 16},
xticklabels = {,,,,,$2^{5}$,$2^{6}$,$2^{7}$,$2^{8}$,$2^{9}$,$2^{10}$,$2^{11}$,$2^{12}$,$2^{13}$,$2^{14}$,$2^{15}$,$2^{16}$},
ymin=0,
ymax = 1.5,
grid=major,
legend pos= south west]

		\addplot[mark= +, smooth ,black, thick] plot coordinates {
			(5, 1.08) 
			(6, 0.98) 
			(7, 0.96)
			(8, 0.99)
			(9, 1.06)
			(10, 1.09)
			(11, 1.28)
			(12, 1.33)
			(13,1.35)
			(14,1.40)
			(15,1.41)
			(16,1.42)
		};
		\addlegendentry{MKRE}

		\addplot[mark = o, smooth, violet, thick] plot coordinates {
			(5, 0.73)
			(6, 0.76) 
			(7, 0.78)
			(8, 0.78)
			(9, 0.82)
			(10, 0.80)
			(11, 0.79)
			(12, 0.85)
			(13, 0.93)
			(14,0.95)
			(15,1.05)
			(16,1.16)

		};
		\addlegendentry{Yggdrasil}
		\addplot[mark=x, smooth,blue, thick] plot coordinates {
			(5, 0.78)
			(6, 0.82) 
			(7, 0.85)
			(8, 0.83)
			(9, 0.80)
			(10, 0.78)
			(11, 0.79)
			(12, 0.76)
			(13,0.81)
			(14,0.82)
			(15,0.85)
			(16,0.93)

		};
		\addlegendentry{\name}

%

		\end{axis}
		\end{tikzpicture}}
	\caption{The total compression rate for different values of \sorg using \name, Yggdrasil~\cite{yggdrasil} and MKRE~\cite{rasmus} and \name for symbol size $\bits=8$. (Lower is better.)}
	\label{fig:result2}
\end{figure}
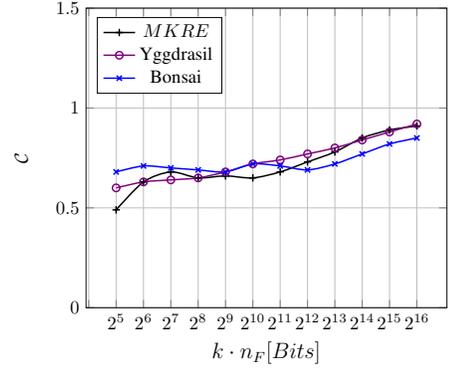
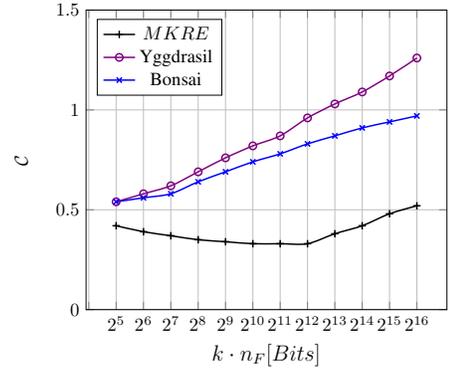
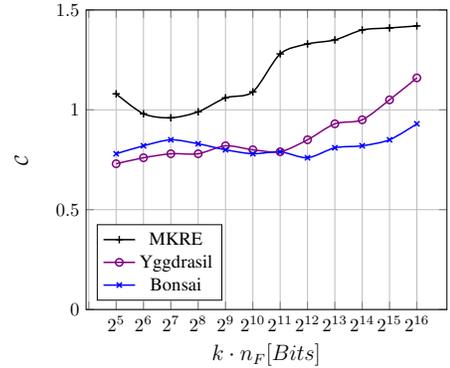

\textbf{Comparison between Bonsai, Yggdrasil and MKRE:}
Fig.~\ref{fig:result2} shows the comparison of \name with the
closest related work, namely
Generalized Deduplication via Multi-Key Revealing Encryption (MKRE)
~\cite{Vestergaard2019b},
and Dual Deduplication via Yggdrasil \cite{yggdrasil}.
In terms of the global compression rates,
as we saw in Fig.~\ref{fig:result1},
\bonsai has a better compression rate for \bits=8
for all datasets used in this work;
therefore, this figure only contains the compression rate
for \bits=8.

In this Fig.~\ref{fig:result2}, we compare the total compression ratio gained for each dataset using different values of \sorg.
As illustrated in this figure, the best compression rate achievable by MKRE~\cite{Vestergaard2019b} and Yggdrasil is nearly the same as the compression ratio achieved by \name for HDFS datasets,
with \name eventually outperforming both Yggdrasil and MKRE
for large files ($\bits\cdot\sorg \geq 2^{12}$).
In more homogeneous datasets, such as DVI, MKRE outperforms both Yggdrasil and \name
by achiveing a compression rate of 0.33 for file size of $2^{10}$,
compared to 0.82 for Yggdrasil and 
0.74 for \name.
Whereas \name outperforms MKRE when using datasets with more diversity,
such as enron-mail dataset.
For example, The difference between compression rate of \name and MKRE 
for a file size of $\bits\cdot\sorg = 2^{9}$
is 0.26.
We note that 
\name provides information-theoretic security
compared to cryptographic security provided by MKRE,
giving an edge to \name in many use cases, while maintaining a competitive
compression potential.

This figure also suggests that in all datasets,
\name outperforms Yggdrasil when the file size is 
$\bits\cdot\sorg \geq 2^{10}$.
In lower chunk sizes, Yggdrasil outperforms \name
in both HDFS logs and enron-mail dataset, yggdrasil outperforms \name
for a file size of $\bits\cdot\sorg \leq 2^{10}$.
However, unlike \name,
Yggdrasil finds potential deduplications by brute-force search,
which introduces huge time complexity for large amounts of data.



\begin{figure}[!t]
	\vspace{-2em}
	\centering
	\begin{tikzpicture}[]
		\begin{axis}[scale=0.7,
			ybar stacked,
			ylabel = {\compratio},
			ymin=0,
			ymax = 1,
			symbolic x coords={
				 $GD$, $ME$, $YG$, $BN$, $BT$
			},
			bar width=8pt,
			xtick=data,
			bar shift=-4pt,
			legend style={legend columns=-1,at={(0.167,0.93),},anchor=west},
			]
			\addplot+[fill=green, on layer={axis foreground}] coordinates {($YG$,0.6312)($GD$,0.8106) ($ME$,0.6508) ($BN$, 0.7118) ($BT$, 0.3517)};
			\addplot+[fill=blue, on layer={axis foreground}] coordinates {($YG$,0.1819)($GD$,0) ($ME$,0.1053) ($BN$, 0.05) ($BT$,0)};
			\legend{\strut \cloud, \strut \user};
		\end{axis}
	\end{tikzpicture}
	\vspace{-1em}
	\caption{Compression ratios for different deduplication techniques on files from DVI dataset for $\sorg=256$ and $\bits=8$. $ME$ and $YG$ denote Multi-key revealing encryption~\cite{rasmus} and Yggdrasil~\cite{yggdrasil} respectively. $BN$ denotes \name. $GD$ and $BT$ denote generalized deduplication~\cite{Vestergaard2019a} and Brotli~\cite{alakuijala2018brotli} respectively. $GD$ and $BT$ do not provide privacy for the users' data.}
	\label{fig:resultcomp}
\end{figure}
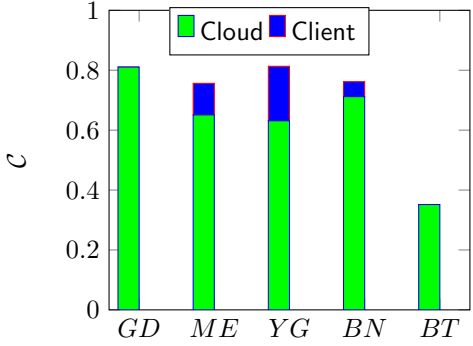

Fig.~\ref{fig:resultcomp} provides a more detailed side-by-side comparison of the
best achievable compression rate on the HDFS dataset using different methods with file size of $\sorg=256$ and symbol size of $\bits = 8$.
In this case, Yggdrasil achieves the best compression rate on the \cloud (0.6312\%),
with the expense of high storage requirements on the \user (0.1819).
MKRE achieves the best total compression rate of 0.7561 between the three methods that provide privacy, i.e., Yggdrasil (0.8131) and \name (0.7668) .
We note that \name holds the middle ground by providing competitive compression rate,
both in terms of total compression rate and the compression rate on the \cloud.
We also note that the privacy provided by \name holds without any cryptographic assumptions.
AS this figure suggests, Brotli outperforms all deduplication schemes
by achieveing a compression rate of 0.3517. However, this is expected as
the DVI dataset is a highly compressable dataset with a low number of duplicate files, which works well with traditional compression techniques as opposed to deduplication schemes.

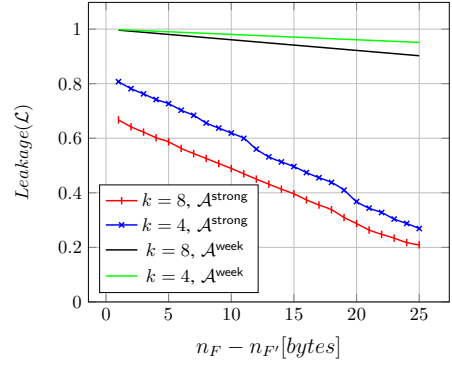
\begin{figure}[!t]
	\centering
	\begin{tikzpicture}
	[scale=0.7]
	\begin{axis}[
	xlabel= {$\sorg - \sbase [bytes]$},
	ylabel=$Leakage(\leakage)$,
	xlabel style={font=\large},
	ymin=0,
	grid=major,
	legend pos= south west]

%

	\addplot[mark = |, smooth, red, thick] plot coordinates {
		(1, .6673) 
		(2, .6417) 
		(3, .6226)
		(4, .6019)
		(5, .5867)
		(6, .5627)
		(7, .5439)
		(8, .5258)
		(9, .5073)
		(10, .4891)
		(11, .4696) 
		(12, .4502) 
		(13, .4317)
		(14, .4132)
		(15, .3963)
		(16, .3741)
		(17, .3553)
		(18, .3381)
		(19, .3097)
		(20, .2873)
		(21, .2638) 
		(22, .2478) 
		(23, .2339)
		(24, .2177)
		(25, .2091)

	};
	\addlegendentry{$\bits=8$, \Sadversary}

	\addplot[mark=x, smooth,blue, thick] plot coordinates {
		(1, .8073) 
		(2, .7817) 
		(3, .7626)
		(4, .7419)
		(5, .7267)
		(6, .7027)
		(7, .6839)
		(8, .6558)
		(9, .6373)
		(10, .6191)
		(11, .5996) 
		(12, .5602) 
		(13, .5317)
		(14, .5132)
		(15, .4963)
		(16, .4741)
		(17, .4553)
		(18, .4381)
		(19, .4097)
		(20, .3677)
		(21, .3438) 
		(22, .3277) 
		(23, .3039)
		(24, .2877)
		(25, .2691)
	};
	\addlegendentry{$\bits=4$, \Sadversary}

	\addplot[
	domain = 1:25,
	thick
	] {(256-x)/256};
		\addlegendentry{$\bits=8$, \Wadversary}

		\addplot[
	domain = 1:25,
	thick,
	green
	] {(512-x)/512};
			\addlegendentry{$\bits=4$, \Wadversary}

	\end{axis}
	\end{tikzpicture}
	\caption{Expected leakage for $\bits\cdot \sorg = 2048\, bits$  for HDFS dataset for different values of $\sbase$ when the PRNG has been broken. Expected value is calculated for 9999 chunks.}
	\label{fig:leakage1}
\end{figure}

\textbf{Privacy:}
Fig.~\ref{fig:leakage1} shows the amount of leakage for a strong adversary and weak adversary when the PRNG has been broken.
This figure suggests that
\name leaks significantly less information to a strong adversary compared to a weak adversary.
As a practical example, for $\bits=8$ and $\sorg-\sbase = 15$,
Leakage to \Sadversary is 0.3963 compared to
0.94 leakage to \Wadversary.
 This is due to the fact that the strong adversary already has a lot of information about the data before receiving the outsourced file \outsource.
We note that in a real-life scenario, the adversary is somewhere between the \Wadversary and the \Sadversary.

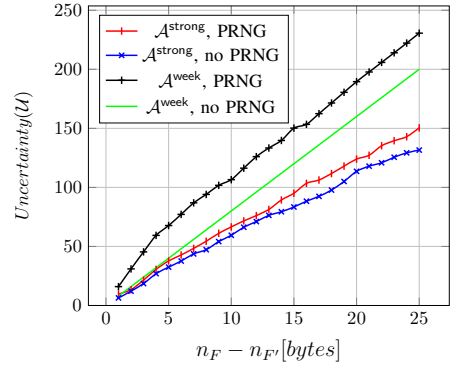
\begin{figure}[!t]
	\centering
	\begin{tikzpicture}
	[scale=0.7]
	\begin{axis}[
	xlabel= {$\sorg - \sbase [bytes]$},
	ylabel=$Uncertainty(\unmetric)$,
	xlabel style={font=\large},
	ymin=0,
	grid=major,
	legend pos= north west]

	%

	\addplot[mark = |, smooth, red, thick] plot coordinates {
	(1, 9.63) 
(2, 13.39) 
(3, 21.45)
(4, 30.59)
(5, 37.76)
(6, 42.77)
(7, 48.36)
(8, 54.40)
(9, 61.11)
(10, 66.51)
(11, 71.62) 
(12, 76.12) 
(13, 81.33)
(14, 89.37)
(15, 95.02)
(16, 103.42)
(17, 106.24)
(18, 111.74)
(19, 118.00)
(20, 123.95)
(21, 127.17) 
(22, 135.25) 
(23, 139.51)
(24, 142.91)
(25, 150.39)

	};
	\addlegendentry{$\Sadversary$, PRNG}

	\addplot[mark=x, smooth,blue, thick] plot coordinates {
			(1, 6.39) 
		(2, 12.13) 
		(3, 18.58)
		(4, 26.95)
		(5, 32.47)
		(6, 37.63)
		(7, 43.69)
		(8, 47.23)
		(9, 54.11)
		(10, 59.45)
		(11, 66.27) 
		(12, 71.04) 
		(13, 76.33)
		(14, 79.37)
		(15, 83.40)
		(16, 88.31)
		(17, 92.42)
		(18, 97.74)
		(19, 105.01)
		(20, 113.59)
		(21, 117.97) 
		(22, 120.83) 
		(23, 125.53)
		(24, 129.26)
		(25, 131.59)
	};
	\addlegendentry{$\Sadversary$, no PRNG}

	\addplot[mark= +, smooth ,black, thick]
	plot coordinates {
	(1, 15.99) 
	(2, 30.98) 
	(3, 45.38)
	(4, 59.36)
	(5, 67.69)
	(6, 77.12)
	(7, 86.93)
	(8, 94.03)
	(9, 101.74)
	(10, 106.54)
	(11, 116.27) 
	(12, 126.04) 
	(13, 133.36)
	(14, 139.73)
	(15, 150.20)
	(16, 153.38)
	(17, 162.42)
	(18, 171.47)
	(19, 180.50)
	(20, 189.53)
	(21, 197.71) 
	(22, 205.83) 
	(23, 213.95)
	(24, 222.19)
	(25, 230.59)
};
	\addlegendentry{$\Wadversary$, PRNG}

	\addplot[
	domain = 1:25,
	thick,
	green
	] {8*(x)};
	\addlegendentry{$\Wadversary$, no PRNG}

	\end{axis}
	\end{tikzpicture}
	\caption{Expected uncertainty in the \cloud for $ \sorg = 256\, bits$ and $\bits = 8$ for HDFS dataset and different values of $\sbase$. Expected value is calculated for 9999 chunks.}
	\label{fig:uncertainty}
\end{figure}

Fig.~\ref{fig:uncertainty} shows the uncertainty metric for the same variables.
This figure includes the uncertainty for $\bits =8$, where \Sadversary and \Wadversary
are considered in two scenarios, where
(a) the adversary has broken the PRNG, and
(b) the adversary cannot break the PRNG in polynomial time.
We choose $\bits = 8$ as a Byte-sized symbol is more practical in real-life scenarios.
Fig.~\ref{fig:uncertainty} shows that even the strong adversary with a broken PRNG faces a large uncertainty to guess the real file \file.
As an example, deleting 15 symbols or 15 Bytes of data
leads to an average of $95.02$ Shannon bits of entropy 
for \Sadversary with negligible probability of breaking the PRNG,
and $83.40$ bits of Shannon entropy for the said \Sadversary if it breaks the PRNG.
These shanon bits of entropy when the \user performs as low as 9 deletions,
are $61.11$ when the adversary can not break the PRNG,
and $54.11$ when the PRNG is broken.
As a standard in information-theoretic literature, 
49 bits of entropy is considered secure against an adversary with
the goal of identifying the original file~\cite{eastlake2005randomness}.
We note that the assumption of breaking the PRNG is a very hard assumption
for the adversary to fulfill (as it entails guessing the seed used by \user to create \outsource).
Our results suggest that \name is still information-theoretically secure 
even when the output of the PRNG is known to the adversary.
This is not surprising, as such information only leaks the deleted positions, but not the deleted values.

\begin{figure}[!t]
	\centering
	\begin{tikzpicture}
	[scale=0.7]
	\begin{axis}[
	xlabel= {\guess},
	ylabel={Outsources \%},
	xlabel style={font=\large},
	xtick = {2,4,6,8,10,12,14,16,18,20,22,24,26,28,30,32,34,36,38,40},
	xticklabels = {$2^{2}$,,,$2^{8}$,,,$2^{14}$,,,$2^{20}$,,,$2^{26}$,,,$2^{32}$,,,$2^{38}$},
	ymin=0,
	grid=major,
	legend pos= south east]

	%

	\addplot[mark = |, smooth, red, thick] plot coordinates {
		(2, 0) 
		(4, 4.19) 
		(6, 10.75)
		(8, 18.31)
		(10, 28.42)
		(12, 39.53)
		(14, 51.72)
		(16, 64.18)
		(18, 73.29)
		(20, 81.51)
		(22, 89.62) 
		(24, 96.18) 
		(26, 99.54)
		(28, 100)
		(30, 100)
		(32, 100)
		(34, 100)
		(36, 100)
		(38, 100)
		(40, 100)
	};
	\addlegendentry{$\Sadversary$, PRNG}

	\addplot[mark=x, smooth,blue, thick] plot coordinates {
		(2, 0) 
(4, 1.81) 
(6, 7.53)
(8, 14.71)
(10, 24.12)
(12, 35.91)
(14, 47.62)
(16, 58.14)
(18, 69.11)
(20, 78.82)
(22, 86.20) 
(24, 93.32) 
(26, 96.41)
(28, 98.28)
(30, 99.17)
(32, 100)
(34, 100)
(36, 100)
(38, 100)
(40, 100)
	};
	\addlegendentry{$\Sadversary$, no PRNG}

	\addplot[mark= +, smooth ,black, thick]
	plot coordinates {
		(2, 0) 
(4, 0) 
(6, 0)
(8, 4.18)
(10, 9.32)
(12, 16.62)
(14, 25.18)
(16, 36.21)
(18, 47.52)
(20, 59.96)
(22, 71.82) 
(24, 81.66) 
(26, 89.52)
(28, 95.31)
(30, 97.26)
(32, 98.90)
(34, 99.42)
(36, 99.78)
(38, 99.99)
(40, 100)
	};
	\addlegendentry{$\Wadversary$, PRNG}

		\addplot[mark= o, smooth ,green, thick]
	plot coordinates {
		(2, 0) 
		(4, 0) 
		(6, 0)
		(8, 0)
		(10, 2.51)
		(12, 6.62)
		(14, 12.53)
		(16, 21.82)
		(18, 33.15)
		(20, 48.28)
		(22, 60.06) 
		(24, 72.41) 
		(26, 81.21)
		(28, 86.73)
		(30, 90.15)
		(32, 93.59)
		(34, 95.71)
		(36, 97.72)
		(38, 98.53)
		(40, 99.58)
	};
	\addlegendentry{$\Wadversary$, no PRNG}
	%
%
%
	\end{axis}
	\end{tikzpicture}
	\caption{Percentage of chunks that \file is among the top \guess pre-iamges of \outsource with highest probability.
	This values are calculated by a monte-carlo test using 9999 chunks,
for $\sorg=256$ and $\bits=8$.}
	\label{fig:guess}
\end{figure}
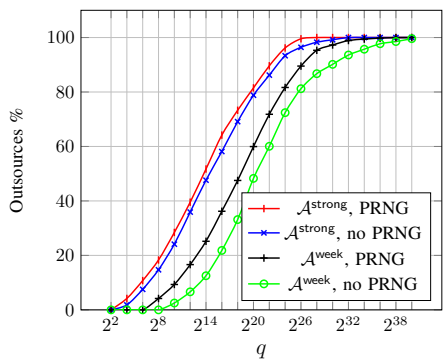

Another interesting metric to analyse the privacy of \name
is the probability of the original \file compared to
other possible pre-images of \outsource.
In order to assign a scientific understanding to this measure,
we calculated the percentage of outsources,
where the original file \file is among the highest probable \guess pre-images
of the outsource \outsource.
In other words, considering the case where the adversary generates all the possible
pre-images of \outsource and sorts then based on their probability,
assuming that the adversary can pick the first \guess chunks in this sorted list,
what is the probability that the original file \file is among the selected pre-images?
We conducted a Monte Carlo test using 9999 chunks
that was created by a client of \name,
where $\sorg = 256$ and $\bits=8$.
Fig.~\ref{fig:guess}
shows the percentage of outsources where the original file is among the
highest probable \guess pre-images of the outsource based on
probability of pre-images.
For the case of \Sadversary
breaking  the PRNG (worst case), Fig.~\ref{fig:guess} shows that
there is less than $50$~\% probability that the original file
is among the top $2^{14}$ pre-images of
the outsource \outsource.
In fact, the strong adversary \Sadversary has to
select the highest $2^{32}$ pre-images
based on their probability to have at least 90\% probability of having the original file
 among the selected pre-images.
 From a practical perspective, selecting from a pool of 4 Billion pre-images for each chunk introduces a high degree of uncertainty for recovering any useful information (although clearly not as strong as cryptographic security).


\vspace{-5pt}
\section{Conclusion}\label{sec:conclusion}
This paper proposes \name, a privacy-aware dual deduplication system that preserve the privacy of \user's data while achieving attractive compression for cloud storage providers. 
Similarly to Yggrasil \cite{yggdrasil}, Bonsai's privacy relies on deletions of symbols from the original data; however, Bonsai achieves significantly better compression rates on the client side by replacing random deletion with deletion on locations indicated by a PRNG. This behavior mimics a deletion channel.
We implement a filtering method using multiple seeds of a PRNG function to ensure that the generated base can be deduplicated in the cloud with higher probability.

To improve its compression capabilities and enable efficient search, the \cloud in \bonsai periodically generates a \policy, which tells to the \users the expected probability distribution of symbols in the outsourced data.
The \user uses the \policy to generate deduplication-friendly data to outsource to the \cloud.
Bonsai implements an innovative mapping function in the \cloud combined with a forest data structure in order to
efficiently find and deduplicate incoming data. 
Our experiments show that \bonsai achieves significant compression rates compared to state-of-the-art (generalized) deduplication approaches. 

In addition, \bonsai protects the confidentiality of \users' data thanks to a deletion-channel-like approach. We argue the proposed approach provides good privacy guarantees: our analysis shows that an attacker faces a high degree of uncertainty when trying to reconstruct \users' original data from what is outsourced. 
 In real-life scenarios, 
The strongest possible adversary faces an average 83.40 Shannon bits of entropy while trying to guess the original data,
even if the adversary has full knowledge about
the characteristics and statistics of the original data source. 
We further show that 
there is only 50\%
chance that the adversary can guess the correct original data
from the received outsourced data,
by checking $2^{15}$ possible strings for each chunk of data.
Even if the adversary has the possbility to check $2^{32}$
possible strings,
The chance of guessing the correct file for each chunk is still less than 90\%.

Future work will study additional mechanisms for Bonsai to improve compression, e.g., using more compact representations for the sorting step, improve privacy, e.g., combine random deletions with value changes or insertions at the \user side, and efficient secure sharing mechanisms with third parties.

\bibliographystyle{IEEEtran}
\bibliography{ref.bib}

\end{document}